\documentclass[aps,prb,twocolumn,superscriptaddress]{revtex4-1}
\usepackage{bm,wrapfig}

\usepackage{graphicx}
\usepackage{epsfig}
\usepackage{amsfonts}
\usepackage{amsmath}
\usepackage{amssymb}
\usepackage{color}

\newcommand\ba{\begin{eqnarray}}
\newcommand\ea{\end{eqnarray}}
\newcommand\be{\begin{equation}}
\newcommand\ee{\end{equation}}

\newcommand{\ct}{\cite}

\def\non{\nonumber}

\def\la{\lambda}

\begin{document}

\title{
Dynamical synchronization transition in interacting electron systems
}
\author{Tanay Nag}
\affiliation{Max Planck Institute for the Physics of Complex Systems, Dresden 01187,
Germany
}

\author{ Robert-Jan Slager}
\affiliation{Max Planck Institute for the Physics of Complex Systems, Dresden 01187,
Germany
}
 
\author{Takuya Higuchi }
\affiliation{Chair of Laser Physics, Department of Physics, Friedrich-Alexander-Universit\"at Erlangen-N\"urnberg (FAU), Staudtstr. 1, D-91058 Erlangen, Germany
}

\author{Takashi Oka }
\affiliation{Max Planck Institute for the Physics of Complex Systems, Dresden 01187,
Germany
}
\affiliation{Max Planck Institute for Chemical Physics of Solids, Dresden 01187,
Germany
}

\maketitle

{\bf Synchronization is a ubiquitous phenomenon in nature
and we propose its new perspective in ultrafast dynamics in interacting electron systems. 
In particular,  using graphene irradiated by an intense bi-circular pulse laser as a prototypical and experimental viable example, 
we theoretically investigate how to selectively generate a coherent 
oscillation of electronic order such as charge density waves (CDW). 
The key is to use tailored fields that match the crystalline symmetry broken by the target order. 
After the pump, a macroscopic number of electrons start 
oscillating and coherence is built up through a transition. 
The resulting physics is detectable as a coherent light emission at the synchronization frequency and may be used as a 
purely electronic way of realizing Floquet states respecting exotic 
space time crystalline symmetries. In the process, we also explore possible flipping of existing static CDW orders and 
generation of higher harmonics.
The general framework for the coherent electronic order is found to be
analogous with the celebrated Kuramoto model, describing the classical synchronization of coupled pendulums.
}

Control of quantum matter using non-equilibrium means is an important topic in fundamental science. 
In ultrafast pump-probe spectroscopy, well-controlled intense lasers are used to 
induce non-equilibrium phase transitions 
(reviewed in Refs.\cite{Iwai2006,Basov11,YONEMITSU20081,Aoki2014,Giannetti2016})
and  
one of the initial ideas was to use the  laser excited carriers to trigger a 
photo-induced insulator-to-metal transition in strongly correlated materials\cite{Cavalleri01,Iwai2003,Perfetti06}. 
In this way, it is able to {\it destroy} orders present in the materials. However, 
{\it creating} a non-trivial order is not straightforward. 
One route is to resonantly pump coherent phonon oscillations optimizing the lattice structure
to favor interesting electronic orders \cite{Kaiser14,Hu14,Mankowsky14,Nova16}. 
Quench induced topological phases would also fall in this  category \cite{PhysRevB.88.104511,PhysRevLett.113.076403}.
Another route is Floquet engineering~\cite{Eckardt2017,Okareview18}, a control of quantum states by periodic driving, 
and nontrivial topological states have been proposed \cite{oka09,lindner11a,Kitagawa10,FloquetMajorana,Rudnerprx} and realized \cite{Wang2013,Jotzu2014,rechtsman13}. 
Here, we study a hybrid of these two ideas. 
Namely, we study how to generate coherent electron oscillations in interacting electron systems with an 
aim to realize novel Floquet states. 
To be more specific, we investigate oscillating orders with frequency $\Omega_0$ 
\be 
\langle c^\dagger_a (t)c_b (t) \rangle = \Delta_{ab}e^{-i(\Omega_0 t-\phi)}+\ldots
 \label{eq_order}
\ee
that did not exist in the groundstate but are created by a short but intense pump laser field. Here,
$a,b$ denote the position (and spin, orbital$\ldots$) indices and the amplitude $\Delta_{ab}$ is at most a slowly changing function.
Various oscillations of density waves (CDW, spin density wave, bond density wave etc.) fit into this category. 
We note that a dynamical order ``Floquet condensation" analogous to (\ref{eq_order}) 
was also found in strongly interacting gauge theories, but the microscopic understanding of the mechanism is still lacking~\cite{Kinoshita2018}.

\begin{figure*}[bht]
\begin{center}
\includegraphics[width=16cm]{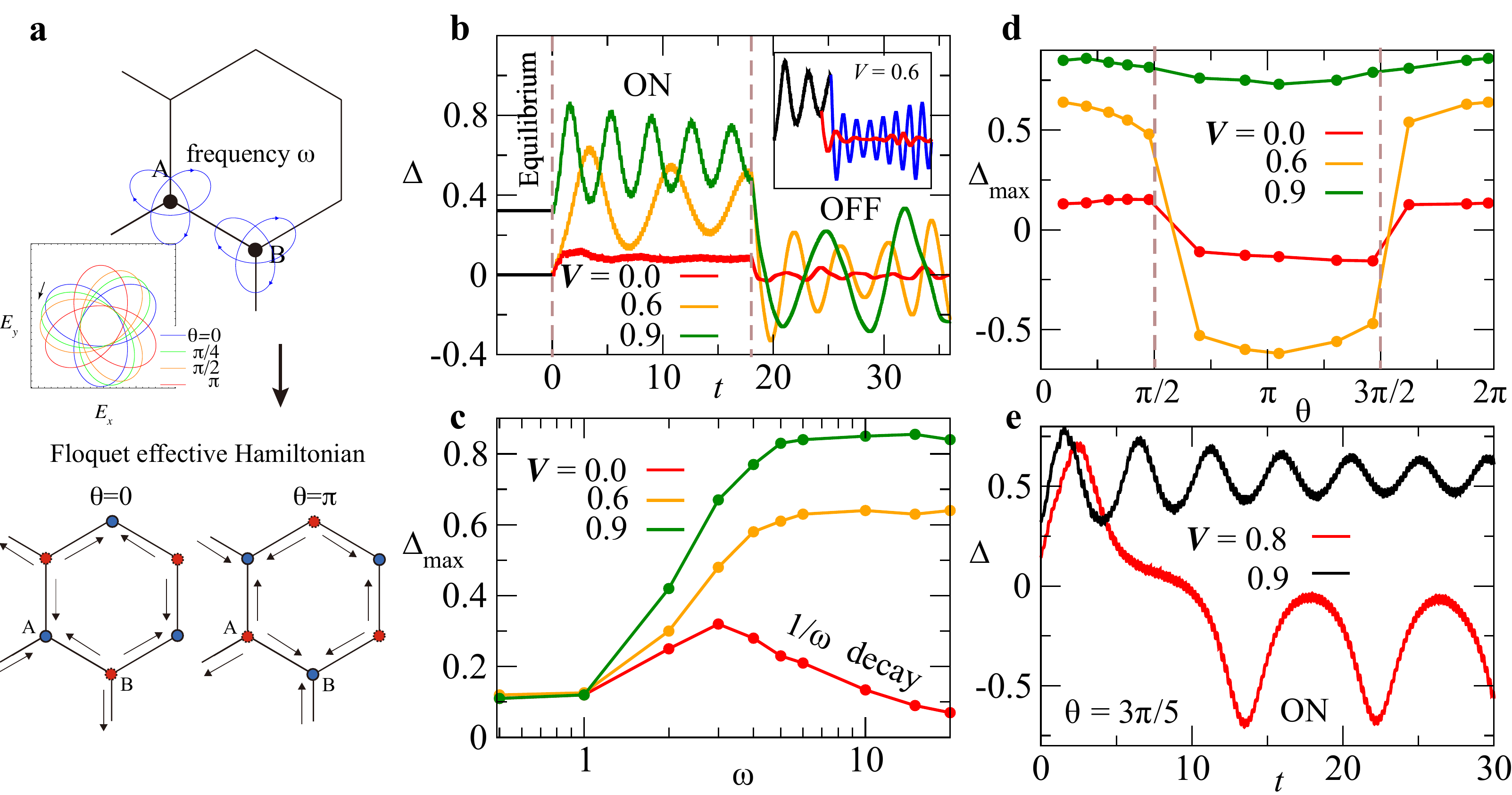} 
\end{center}
\caption{{
{\bf Dynamics of CDW oscillation.}
\textbf{a},  Bi-circular laser field, as depicted by the blue curve, makes the A and B sublattices in the honeycomb lattice are inequivalent 
(inset: angle dependence). When it is ON, effective terms 
such as an on-site AB-sublattice alternating potential is dynamically induced, which triggers a charge transfer (arrows). }
\textbf{b},  Time evolution of the CDW order parameter. 
The oscillation amplitude and the mean values of $\Delta$
are both enhanced by the interactions. 
Inset shows that for a given value of interaction strength ($V=0.6$), the relaxation dynamics bears a prominent memory effect depending on the values of CDW order particularly at 
instant when laser field is switched off. ($\omega=10.5$, $\theta=\pi/10$).
\textbf{c},  The maximum value of the CDW order $\Delta_{\rm max}$, during the bi-circular driving, shows a 
significant increase  in the interacting case for $\omega>2$. ($\theta=\pi/10$). 
\textbf{d},  The dependence of $\Delta_{\rm max}$ on the angle $\theta$ of the bi-circular field. 
It switches sign around $\theta=\pi/2$ and $\theta=3\pi/2$ ($\omega=10.5$).
\textbf{e}, 
Starting within the CDW phase,  it is possible to  flip the order when we are near both the phase boundary and $\theta=\pi/2$. 
} \label{figCDW} 
\end{figure*} 

A desired coherent oscillation can be selectively created by using  pump fields that are carefully tailored. 
This is in contrast to the Higgs amplitude mode
\cite{Littlewood81,Littlewood82,Matsunaga2011,Matsunaga2014,Tsang76, Demsar99,PekkerHiggsreview}, {\em i.e.} a
small oscillation of an already existing groundstate order, which can be 
triggered by a delicate but featureless excitation across the gap. 
To make the argument solid, we use light irradiated graphene \cite{graphene} as an example since 
it is a prime testbed for theoretical and experimental researches \ct{brumer86, oka09, ishikawa10, 
 rioux11, kelardeh15, higuchi17}.
We study electrons on the honeycomb lattice at half-filling described by the Hamiltonian 
\be
H = \frac{1}{N}\sum_{{\bm k}\sigma} {\hat \psi}^\dagger_{{\bm k}\sigma} {\cal H}({\bm k}+{\bm A}) {\hat \psi}_{{\bm k}\sigma},
{\cal H}({\bm k}) = \begin{pmatrix}
0 & h({\bm k} ) \\
h({\bm k} )^*  & 0
\end{pmatrix}
\label{eq_hma}
\ee
with ${\hat \psi}^\dagger_{{\bm k}\sigma}=(c^\dagger_{A{\bm k}\sigma},c^\dagger_{B{\bm k}\sigma})$ in the momentum ${\bm k}$ space, 
while $A$, $B$ are the sublattice indices.  $h({\bm k} ) =\sum_{l=0}^2J e^{-i {\bm k} \cdot {\bm e}_l}$, ${\bm e}_l =  (\cos\phi_l, \sin\phi_l)$ with $\phi_l= \pi/2 + 2\pi l/3$ sets the hopping and $N$ is the number of unit cells in the lattice. 
The CDW and complex bond order parameters are
\be
\Delta=\frac{1}{N}\sum_{{\bm k}}\Delta_{\bm k}, \; b^\pm=\frac{1}{N}\sum_{{\bm k}}b^\pm_{\bm k}
\ee
with $\Delta_{\bm k}=\sum_\sigma\langle[c^\dagger_{A{\bm k}\sigma}c_{A{\bm k}\sigma}-c^\dagger_{B{\bm k}\sigma}c_{B{\bm k}\sigma}] \rangle$
and $b^{+(-)}_{\bm k}=\sum_\sigma\langle c^\dagger_{A(B){\bm k}\sigma}c_{B(A){\bm k}\sigma} \rangle$. {
How can we dynamically induce a CDW oscillation using laser? }
The order breaks the sublattice symmetry, reducing the lattice's $C_{6v}$ crystalline symmetry down to $C_{3v}$. 
Bi-circular laser, which is itself an experimentally well-established technique~\cite{Kfir:wd}, 
can be expressed by a gauge field ($A=A_x+iA_y$) 
\be
A=A_Le^{i\omega t}+A_Re^{-2i\omega t+i\theta}
\label{eq_bcf}
\ee
also has this lower $C_{3v}$ symmetry (we set $A_R=A_L$). 
Under the electric field $E(t)=-{\partial_t A}$, 
the A and B sublattices become inequivalent (Fig.~\ref{figCDW}\textbf{a}) and 
is expected to trigger the desired oscillation of $\Delta$. 
{ A bi-circular laser effectively induces  terms that break the $C_{6v}$ symmetry and the
Floquet effective Hamiltonian includes
a AB-sublattice potential $m_{\rm eff}\sum_{{\bm k}\sigma}
[c^\dagger_{A{\bm k}\sigma}c_{A{\bm k}\sigma}-c^\dagger_{B{\bm k}\sigma}c_{B{\bm k}\sigma}]$
(see supplementary II B).
This can push the electrons from A to B sublattices (or vise versa) initiating the oscillations. 
}

Collective dynamics occur when we add electron-electron interactions
\be
H_{int}=\frac{U}{2}\sum_i n_{i} n_{i}+ V\sum_{\langle i,j\rangle}n_{i}n_{j}
\label{eq_int2}
\ee
to the free Hamiltonian~(\ref{eq_hma}), where $U$ and $V$ denote the on-site and nearest neighbor Coulomb repulsion respectively ($n_i=\sum_\sigma c^\dagger_{i\sigma}c_{i\sigma}$, $\langle i,j\rangle$: 
nearest neighbor pairs). 
The real time dynamics of the extended Hubbard model at half-filling on the honeycomb lattice
is studied within the time-dependent mean-field approximation using 
\begin{eqnarray}
{\cal H}_{\rm MF}({\bm k})&=&  \frac{1}{2}\left[ \begin{array}{cc} 
(\frac{U}{2} - 3 V) \Delta & 0 \\
0 & -(\frac{U}{2}- 3 V) \Delta
\end{array} \right]\non\\
&+& {\cal H}({\bm k})+\bigl(\frac{U}{4}-\frac{3V}{2}\bigr)I
\label{eq_ham_mf1}
\end{eqnarray}
to govern the time evolution. 
Below, we set $J=1$ to fix the energy scale and  use $A_R=A_L=1$. 
Since  the combination $U/2-3V$ is the only relevant 
interaction parameter within this approximation, 
we set $U=0$ and use $V$ as the parameter representing the correlation effect.
Within the mean field approximation, the critical value for the groundstate CDW transition is $V_c=0.78$.
In graphene \cite{graphene}, the material parameters are known to be close but below the critical value accommodating a semi-metallic phase~\cite{wehling11}.


\vspace{1em}
\noindent 
{\bf \center CDW oscillation and flipping}\newline
Figure~\ref{figCDW}\textbf{b}  shows the time evolution of the CDW order triggered by the bi-circular pulse field (\ref{eq_bcf}).
The field is suddenly ramped up and down at the beginning and end of the ON region. 
We checked that the findings qualitatively remain unaltered for slow ramping protocols. 
Interactions substantially affect the CDW order. They give rise to a pronounced 
oscillation {both in the ON and OFF region,} where the amplitudes are controlled by the new energy scale set by $V$. 
Inset shows that even for the same interaction and pulse strength, 
the oscillation amplitude of $\Delta$ during the relaxation dynamics
significantly depends on the value of $\Delta$ at which the field is switched off. 
 This memory effect (or initial value effect) is completely absent for the non-interacting case.
In Fig.~\ref{figCDW}\textbf{c}, we further investigate
 $\omega$ and $V$ dependence of $\Delta_{\rm max}$.
The first observation is that CDW oscillations can be induced in a wide range of 
pump frequencies. However, a strong upturn occurs when 
 $\omega=2$ is exceeded. This is also where the interaction starts to assist the oscillation. 
For higher $\omega$, we find that $\Delta_{\rm max}$ saturates for the interacting case,
in contrast to the non-interacting case, where $\Delta_{\rm max}$ falls off as $1/\omega$, 
following high frequency perturbation theory (see Supplementary II).

\begin{figure*}[ht]
\begin{center}
\includegraphics[width=16.cm]{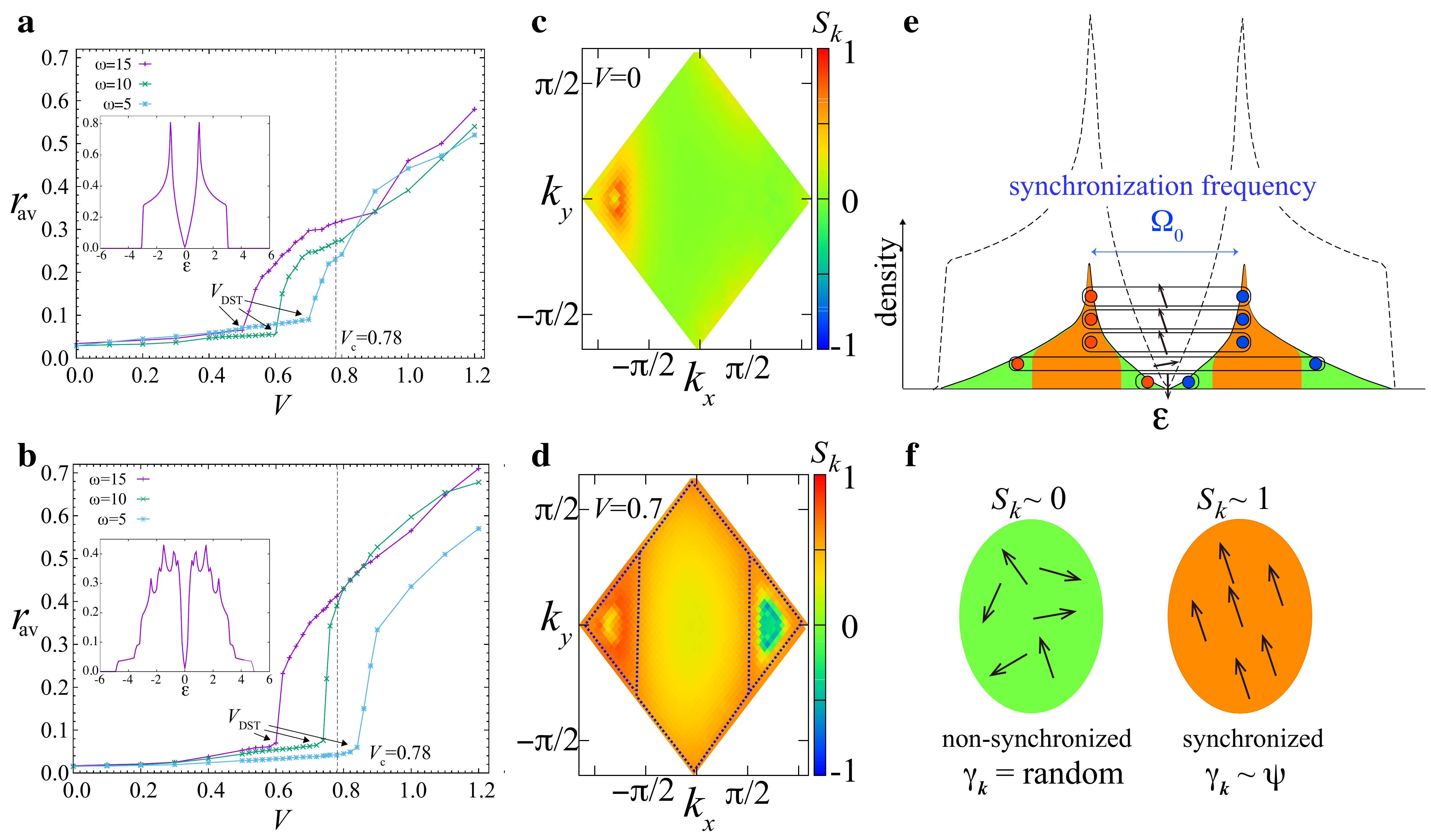}
\end{center}
\caption{
{\bf Dynamical synchronization transition.}
\textbf{a},   The synchronization order parameter (average value) $r_{\rm av}$ for the model on a honeycomb lattice ($\theta=\pi/10,;N=100\times 100$).  Inset shows the density of states. 
$V_{\rm DST}$ is the critical interaction strength for the DST to occur. 
\textbf{b},  Same as \textbf{a} but with an altered lattice model having a reduced singularity in its density of states. 
\textbf{c},  
\textbf{d},  
Momentum resolved synchronization correlation  $S_{\bm k}$ for $V=0.0$ and $V=0.7$, respectively. ($\omega=10$, $\theta=\pi/10$). 
Blue dotted line in \textbf{d} shows the contour $|h({\bm k})|=1$ signifying the van-Hove singularity.  \textbf{e},  
 Schematic plot of the electronic density of states  (dashed) 
in the honeycomb lattice and the distribution 
of electron-hole pairs (solid). 
The latter is closely connected to $\int d^2{\bm k} |\Delta_{\bm k}+iJ_{\bm k}|
\delta(\varepsilon-|h({\bm k})|)$
and quantifies excitations 
due to pumping in the presence of interaction. 
The blue and red circles represent electron-hole pairs and the arrow refers to their polarization direction $\gamma_{\bm k}$. 
\textbf{f},  
Green and red regions schematically depict the 
non-synchronized and synchronized states respectively 
in momentum (\textbf{c},  \textbf{d}) and energy (\textbf{e}) spaces. 
The arrows illustrate $\gamma_{\bm k}$ and in the 
synchronized state,  they rotate with a synchronization frequency $\Omega_0$
and its direction defines the collective phase
 $\gamma_{\bm k}\sim \psi$. 
} \label{figSync}
 \end{figure*}

The angle $\theta$ of the applied bi-cicular field plays a crucial role in controlling the CDW order (see Fig.~\ref{figCDW}d). 
For  $\theta=0,\;\pi$, it induces an effective AB-sublattice potential that favors a CDW order,
and $\Delta_{\rm max}$ becomes largest 
with opposite signs as shown in Fig.~\ref{figCDW}\textbf{d}. 
Flipping of already existing CDW order is possible (Fig.~\ref{figCDW}e).
We find that $\Delta$ 
can be flipped when we start in the ordered phase but not so far from the phase boundary. 
As for the direction of the bi-circular field,
the flipping started to occur above $\theta=\pi/2$ and became maximum around $3\pi/5$.  
Although we don't have a clear understanding of the flipping mechanism yet, 
at least in the non-interacting case, 
we find an analogous picture to the spin-echo ($\pi$-pulse) technique  in NMR. 
Specifically, if we view the effective Floquet Hamiltonian (see supplementary II) as a 
spin Hamiltonian $\mu {\bf B}_{\boldsymbol k}\cdot {\boldsymbol \sigma}$, the order parameter 
$\Delta$ is analogous to a $z$-direction magnetization. 
The situations for $\theta=0,\pi$ correspond to having a $z$-field
since an effective AB-sublattice potential is induced and tries to align the spin 
in the $z$-direction. However, once the order is established, a 
$z$-magnetic field that commutes with the order is not able to change it. 
On the other hand, the $\theta=\pi/2$ situation corresponds to having a
transverse field, and it can be used to rotate the existing order $\Delta$
by suitably tuning $\sigma_x$ and $\sigma_y$ components in the effective Hamiltonian
({see supplementary II}).

\vspace{1em}
\noindent 
{\bf Synchronization order parameter}\newline 
Density oscillations are usually quickly  damped due to dephasing
because electron-hole pair excitations, created by short and intense fields,
spread out broadly in the energy-frequency space \cite{Oka12}. 
{\it Dynamical synchronization transition} (DST),  ubiquitous in non-linear dynamical systems, 
is a mechanism that acts against dephasing~\cite{Kuramoto_book,Acebron05}. 
An ensemble of interacting oscillators with different 
frequencies can oscillate collectively at a single synchronization frequency $\Omega_0$. 
In the present system, the momentum-resolved CDW, current, and bond 
order parameters
($\{\Delta_{\bm k},\; J_{\bm k}=-i(b^+_{\bm k}-b^-_{\bm k}),\; K_{\bm k}=(b^+_{\bm k}+b^-_{\bm k})\}$)
define a three dimensional vector field in the Brillioun zone. 
We find below that, while $K_{\bm k}$ changes only slowly, 
the $(\Delta_{\bm k},J_{\bm k})$ component rotates around the $K_{\bm k}$-axis swiftly
(with natural frequency $\Omega= 2|h({\bm k})|$ in the non-interacting case) and 
plays the role of the angle in the classical oscillator  synchronization problem. 
The angle $\gamma_{\bm k}$ specifying the $(\Delta_{\bm k},J_{\bm k})$ direction
can be thought of as a generalized polarization direction. 
We can define~\cite{Kuramoto_book,Acebron05} the  collective phase $\psi$ and amplitude  $r$ by 
 \be
re^{i\psi}=\frac{1}{N}\sum_{{\bm k}}e^{i\gamma_{{\bm k}}},  \;e^{i\gamma_{\bm k}}=-\frac{\Delta_{\bm k}+iJ_{\bm k}}{|\Delta_{\bm k}+iJ_{\bm k}|}.
\label{eq_defr}
\ee
The amplitude $ r\in [0,1]$ plays the role of the synchronization order parameter, and gives the fraction of the synchronized oscillators. 
For example, $r=1$ ($r=0$) 
means full (no) synchronization is achieved. 
A non-zero $r$  signals a transition  from a dephased state to a coherent one. 

Direct evidence for DST can be obtained by evaluating the average value synchronization order parameter $r_{\rm av}$ 
as shown in Fig.~\ref{figSync}\textbf{a}, where we do the time average in the OFF region for a 
considerably long time span  ($15<t<200$). 
As the interaction is increased, the synchronization order parameter 
increases monotonically,  
and we find a singular upturn associated with a non-analytic behavior signifying a collective transition.  
This transition is dynamical in the sense that it depends on how the CDW order
is induced. For example, 
the critical interaction strength,  referred as $V_{\rm DST}$, depends on the 
details of driving i.e., 
the frequency $\omega$ and field strength $A_L=A_R$ of the bi-circular field 
as well as details of the underlying lattice model.
An interesting observation is that $r_{\rm av}$ is non-zero even for the non-interacting case. 
This is due to the van Hove singularity in the density of states (DOS) 
at $\varepsilon_{\rm vH}=\pm 1$ for the honeycomb lattice,
representing a large fraction of coherent electron-hole pairs at
$\Omega_0=2=2|\varepsilon_{\rm vH}|$. 
In order to isolate this effect, 
we study the DST   by using a 
modified lattice Hamiltonian with smoothened and broadened DOS without a singularity (see Fig.~\ref{figSync}\textbf{b}). 
There we find that $r_{\rm av}$ in the non-synchronized phase stays at a smaller value approaching zero 
under a scaling analysis (as $N^{-1/2}$), and the transition becomes much more sharp and pronounced (see supplementary I).

We can obtain a  deeper understanding of DST by 
looking into momentum resolved synchronization correlation defined by ($t_i=15,\;t_f=50$)
\be
S_{\bm k}=\frac{1}{\delta t}\int_{t_i}^{t_f} \Delta(t) \Delta_{\bm k}(t) dt
\label{eq_SC}
\ee
plotted in Figs.~\ref{figSync}\textbf{c,~d}. 
While the correlation is very weak in the non-interacting case,
we see significant correlation built up in the interacting case 
in a red patch around the equal energy ($2|h({\bm k})|=\mbox{const.}$)-contours close to the 
van-Hove singularity (blue dashed contour in Fig.~\ref{figSync}\textbf{d}). 
The synchronization of the momentum modes can thus be depicted as the red region in Figs.~\ref{figSync}~\textbf{f} by 
rotors rotating with a common phase $\gamma_k\sim \psi$, while green region illustrates the 
incoherent non-synchronized case. 
Physically, 
the momentum resolved order $(\Delta_{\bm k},J_{\bm k})$  
is associated to electron-hole pair excitation 
with energy $E_e-E_h\simeq 2|h({\bm k})|$. 
Then moving to the energy $\varepsilon$ space, 
we can think of an intuitive picture as shown in Figs.~\ref{figSync}\textbf{e}. 
The pulse field excites electron-hole pairs with a broad distribution in 
energy space. 
Their polarization (designated by an arrow)
start to rotate individually with frequencies $\simeq 2|h({\bm k})|$. 
Below the DST transition $(V<V_{\rm DST})$, they 
keep rotating incoherently and average out to zero. 
While above the DST $(V>V_{\rm DST})$, coherence among the electron-hole pairs is established collectively through interaction. 
The degree of synchronization 
is increased when the electron-hole pair density becomes higher. 
The synchronization frequency $\Omega_0$ depends on the excitation distribution 
and its value is close to peak of the above distribution. 
Even in the synchronized state, there can be non-synchronized excitations 
associated with energies away from $\hbar \Omega_0$.

The mathematical structure behind the DST 
can be elucidated as  we can relate 
the time evolution within the mean field approximation (\ref{eq_ham_mf1}) 
to a variant of the Kuramoto model \cite{Kuramoto_book,Acebron05} that governs the dynamics of $\gamma_{\bm k}$ 
(supplementary I). 
The mean field coupling can be mapped to the synchronization force and  the laser pumping to an external force. 
This framework is general and applies not only to graphene but to 
wide variety of correlated electron systems even without the van-Hove singularity and to their collective dynamics.

\begin{figure}[tbh]
\begin{center}
\includegraphics[width=8.5cm]{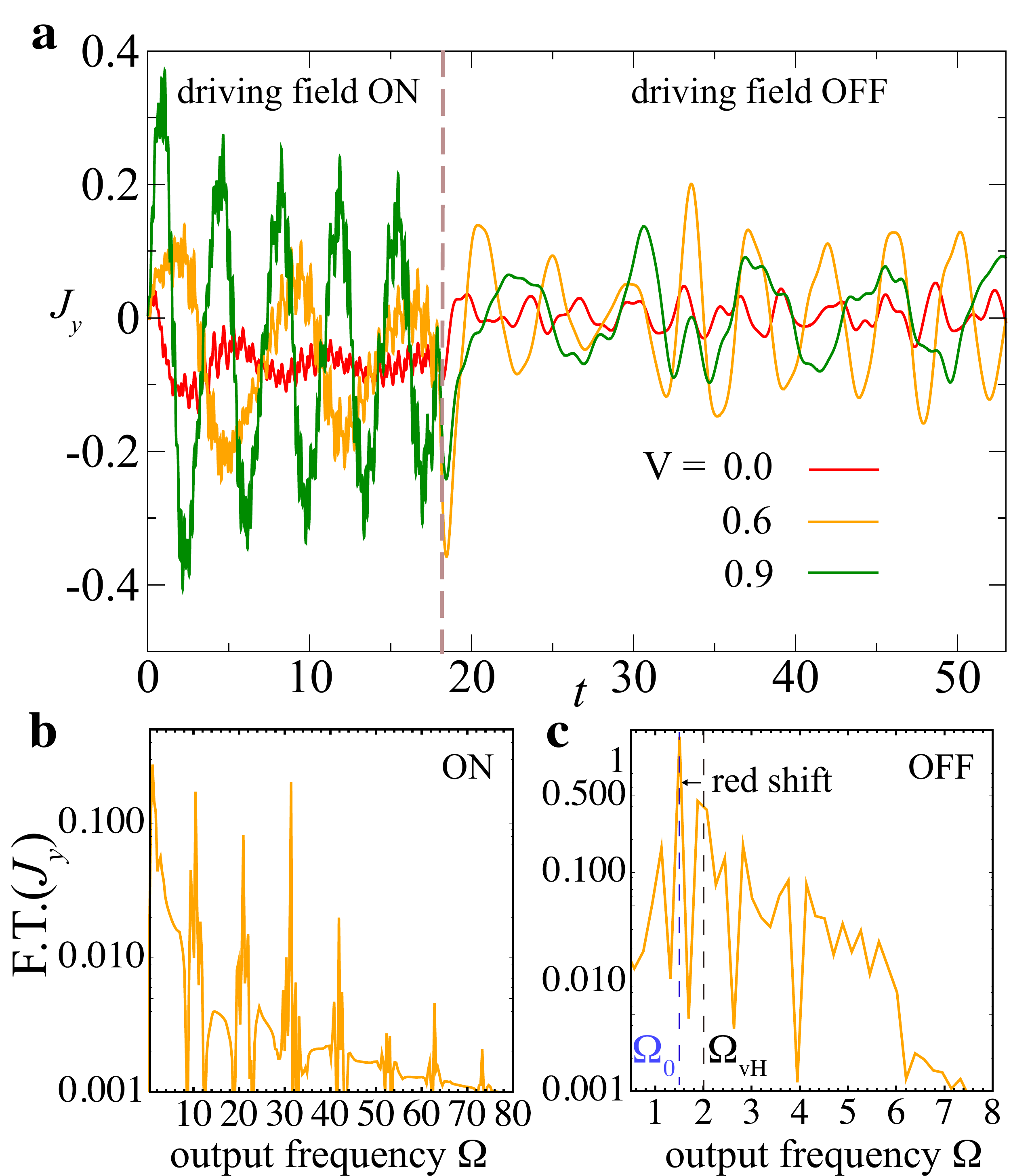} 
\end{center}
\caption{
{\bf Possible experimental observation in the current.}
\textbf{a} The time evolution of the current for the same parameters as Fig.~(\ref{figCDW}\textbf{b}). 
The Fourier transform of $J_y$ in the (\textbf{b}) ON and (\textbf{c}) OFF  regions.  
\textbf{b}  Generation of higher harmonics is observed in the ON region.
\textbf{c} In the OFF region, the synchronization frequency of the oscillation shows a red shift 
from the van Hove singularity  ($\Omega_{\rm vH}=2$) due to correlation effect to $\Omega_0\simeq 1.5$. 
For \textbf{b}  and \textbf{c}  $V=0.6$ is used, while  $\theta=\pi/10,\;\omega=10.5$ is used for all plots.
} \label{figHHG}
 \end{figure}

\vspace{1em}
\noindent 
{\bf Coherent light emission and HHG}\newline
How can we experimentally verify the DST? 
While the CDW is not easily accessible, currents induced by a laser can be measured directly
 as shown by recent work \cite{Yoshikawa736,higuchi17} on high harmonic generation (HHG) in graphene.
 Accordingly, we  now investigate the non-equilibrium evolution of the induced current (${J}_a=\sum_{{\bm k}\sigma}\psi^\dagger_{{\bm k}\sigma}\left[\partial{\cal H}({\bm k}+{\bm A})/\partial k_a\right]\psi_{{\bm k}\sigma}$)
as depicted in Fig.~\ref{figHHG}\textbf{a}. 
The current shows a prominent beating  pattern with increasing $V$ in both ON and OFF regions. 
This beating reflects the CDW oscillation already presented in Fig.\ref{figCDW}, as the current reflects the 
 movement of charge  between the A and B sublattices. 
Most interestingly,  in the ON region,  we observe the generation of higher harmonics i.e. output frequency $\Omega= n\omega$
while investigating the  Fourier component of the current (see Fig.~\ref{figHHG}\textbf{b}).

Concentrating on the OFF region, an
oscillation with more prominent coherence in 
$J_y$ is observed for $V$  near $V_c$. 
A large peak at $\Omega_0\sim 1.5$ is present in the Fourier spectrum of $J_y$ (Fig.~\ref{figHHG}\textbf{c}). 
The synchronization frequency $\Omega_0$ is selected dynamically through a correlation effect
and depends strongly on the electron-hole pair population density as we discussed after Fig.~\ref{figSync}\textbf{e}. 
The synchronization frequency $\Omega_0$ is red shifted from the van-Hove singularity $\Omega_{\rm vH}=2$. 
This is because the electron-electron repulsion $V$, 
which is attractive between electrons and holes, 
reduces the pair energy from the bare value $2|h({\bm k})|$ by its binding energy. 
There has been a 
optical luminescence experiment in graphene and
the luminescence peak is found at 4.62eV which is
reduced from the expected bare  van-Hove energy 5.2 eV \cite{Mak11}. 



\vspace{1em}
\noindent 
 {\bf Road to self-maintained Floquet crystalline states}\newline
What is the consequence of having an oscillating mean field? 
An important point is that  Floquet states are now self-maintained by the electronic degrees of freedom. 
They can remain even after the pump is finished as long as the oscillation exists. 
The Floquet state realized by DST has an  crystalline symmetry now elevated to a space-time crystalline symmetry. 
For example, an oscillating CDW $\Delta=\Delta_0\sin\Omega_0 t$ realizes 
a state with a time-glide symmetry. The mirror symmetry $M$ 
for the mirror plane along the $x$-axis between the AB sites
is broken by $\Delta$, 
but a time-glide symmetry  $H(t, \mathbf{r})=MH(t+T/2, \mathbf{r})M^{-1}$ ($T=2\pi/\Omega_0$) is satisfied.
To make this more specific, we considered the generation of such a naturally induced time glide symmetry and its topological aspects in the Floquet condensation picture explicitly (see Supplementary III).
 Indeed, by reverting to the minimal model introduced above, introduction of an oscillatory CDW term in the Hamiltonian can be shown to result in special edge states specific to Floquet states~\cite{Rudnerprx} for different parameter regimes. 
 These edge states  at the $\pi$-gap in the quasi energy spectrum persist until the gap is closed and are related to the chiral symmetry of the system \cite{Tg1}.  Nonetheless, while this 
 is particular example of a topological symmetry protected state, this mechanism in general opens the door to access many new topological phases protected by new symmetries. 
Parallel to how the concept of topological insulators \cite{Rmp1,Rmp2}
is upgraded to topological crystalline insulators \cite{Clas1,Clas2,Clas3,Clas4,Clas5},
the notion of Floquet topological insulators \cite{oka09,lindner11a}
can be refined by considering the  space-time crystalline symmetry \cite{Xu17}. 
The DST, with the above example in particular, thus provides an experimentally viable route to realize long lived 
Floquet topological crystalline insulators \cite{futurework}.

Another important point is the possibility of phason modes $\phi$ in Eq.~(\ref{eq_order}). 
Windings can exist due to the equivalence $\phi({\bm x})\sim \phi({\bm x})+2\pi$, leading to 
 topological excitations. Examples include vortices and domain walls, which
should bind non-trivial defect or edge modes similar to their static counterparts in Floquet topological
crystalline insulators \cite{Clas6, modesT}. 
We will report on these issues elsewhere. 

\vspace{1em}
\noindent 
{\bf Beyond mean field}\newline
Before closing, let us comment on the validity of our prediction which is 
based on the time dependent mean field approximation. 
In real materials, relaxation due to phonons as well as electron-electron scattering processes 
are important. In particular, in graphene, it is known 
that phonon emission is the dominant relaxation channel, and the typical 
time scale for this to happen is few hundred femtoseconds (fs)~\cite{Gierz:tg}. 
On the other hand, in our simulations, the unit of time is $1.8$ fs considering 
the tunneling amplitude $J=2.3\;\mbox{eV}$ for graphene.
Thus, the synchronization oscillation period $T=2\pi/\Omega_0\simeq 8 \;\mbox{fs}$ is well shorter than the 
relaxation time scale. 
As for electron-electron scattering process, considering the effect of a momentum-time dependent random mass term 
that mimics scattering, we find that the coherence persists up to a certain realistic 
disorder strength (supplementary IV). 
In addition, heating may take place in the ON region. 
However, our protocol uses short pulse excitation and 
we think that heating is not a sever problem in observing DST.
This is because it is known that there exists a Floquet-prethermalization time scale below which the heating 
is negligibly small \cite{PhysRevB.95.014112,Weidinger:tb}.

\vspace{1em}
\noindent 
{\bf Conclusion}\newline
There exists a plethora of studies on Floquet topological insulators in the condensed matter community, but 
up to now, the exciting topological features have been limited in time by the duration of laser pulse. 
We introduced the concept of DST in a simple setup, where a tailored laser pulse is used
to induce a collective electronic order dynamically. 
We find that DST paves the way to realize a Floquet topological state 
that is self-maintained and may become considerably long lived. 
In addition, DST gives a microscopic foundation
to Floquet condensation Eq.~(1) as 
discussed in a holographic Floquet system~\cite{Kinoshita2018}.
Our findings also include general measurable features such as 
coherent light emission and HHG, as well as a 
realistic protocol to flip and control CDW orders.

\newpage
\appendix
\clearpage

\begin{center}
 {\bf Appendix } 
 \end{center}

 In this Appendix we comment on the numerical details underlying the results of the main text. Moreover, we
reveal the analogy between Kuramoto model and the graphene under bi-circular laser field. In particular, we show that the 
two level problem can be mapped to an effective Kuramoto model which qualitatively describes the synchronization phenomena 
occurring in charge density wave order in the presence of interactions. We also compare these results with the non-interacting case. 
Simultaneously, we try to motivate a more physical picture of the synchronization process
using charge density wave order, current order 
and bond order. We then consider another  honeycomb lattice model with an additional hopping term to investigate the 
dynamical synchronization transition (DST) more critically. We find that singularity in the density of states is not a
necessary condition to have DST. 
We extend our results in the context of Floquet perturbation theory to show that in the high frequency limit an inversion 
symmetry breaking mass term indeed generated. We also study the effective hopping terms  individually to understand 
the dyamical behavior of charge density wave order in an extensive way.
After that, we use the effective model to illustrate the emergence of edge states in case of an 
oscillating CDW term that generates a time glide symmetry. This shows that in a Floquet condensation picture, DST setups can 
be used to access new symmetry protected topological phases. Finally, we consider a random momentum-time mass term to 
study the scattering effect on the synchronization where we show that substantial disorder can destroy the 
DST. We further comment on the heating mechanism during the driving and its impact on DST.  

\section{Numerical Details}
We here would like to mention a few technical detail
to calculate the CDW order in presence of interaction.
We consider the momentum space  $2$-level graphene Hamiltonian  (\ref{eq_hma})
to carryout the time-dependent mean field treatment.
First, we determine the initial value of $\rho_A$ and $\rho_B$, before
going into dynamics, by an iterative method for a given
value of $U$ and $V$; in order to this, we consider some
random initial guess for $\rho_A$ and $\rho_B$.
Once $\Delta$ is set in such a way, 
we proceed with time dependent 
Schr\"odinger equation,
governed by the Hamiltonian ${\cal H}_{\rm MF}({\bm k})$ (Eq. 
\ref{eq_ham_mf1})  and solved it again iteratively using the forth order Runge-Kutta method. In the switch ON region with an explicit time dependence Hamiltonian,
we continue the iterative method  replacing 
 $\rho_A$ and $\rho_B$ at
every steps of time with their updated values  as mean field results in a dynamical
evolution of $\rho_A$ and $\rho_B$. In the switch OFF region,
the same process continued without the explicit time dependence of
the Hamiltonian.

\vspace{3em}
\section{On the dynamical synchronization transition}
\label{s1}
\subsection{Time dependent mean field approximation and a link to Kuramoto physics}
\label{sec:modelana}
Here, we outline the connection between the dynamics, based on the time dependent mean field approximation, 
and the Kuramoto model, describing the synchronization transition for coupled pendulum. 
In order to simplify the argument, let us incorporate the effect of the bi-circular laser field 
by an effective AB-sublattice alternating potential $m_A=m_0,\; m_B=-m_0$, which is justified using the 
high frequency expansion (supplementary II). 
The model Hamiltonian then reads 
\be
 H_k(t)= \left[ \begin{array}{cc} 
\lambda \Delta(t) + m_0  & h_{\bm k} \\
h_{\bm k}^* & -\lambda \Delta(t) -m_0 
\end{array} \right]
\ee
with the mean field potentials $\lambda \Delta(t)$ and $\Delta(t)\equiv \sum_{\bm k}\rho_{\bm k}^A(t)-\rho_{\bm k}^B(t)$, where $\lambda=1/2
(U/2-3V)$.
Using  unitary transformations 
$U_1=\exp(-i\sigma_z \phi_k/2)$, $U_2=\exp(-i\sigma_y \pi/4)$ with
$\phi_k={\rm Arctan}({\rm Im}(h_k)/{\rm Re}(h_k))$,  we re-express the Hamiltonian in a more convenient rotated basis~\ct{barankov04}
\be
\hat H_k(t) = U_2 U_1 H_k(t) U_1^{\dagger} U_2^{\dagger}
 = \left[ \begin{array}{cc} 
   |h_k| & \lambda \Delta(t) + m_0\\
 -\lambda \Delta(t) -m_0 & -|h_k| 
\end{array} \right]. 
\ee
In this basis, 
the charge density order becomes $\Delta(t)=-2\sum_k {\rm Re}(u_k^* v_k)$, where
$u_k(v_k)=1/\sqrt{2}(+(-) \exp(-i \phi/2) \psi_k^A + \exp(i \phi/2) \psi_k^B)$. 
The mean field equation of motion governing the dynamics can be recasted
as follows 
\be
i\dot w_k=2 |h_k| w_k +(\lambda \Delta(t)+m_0)(w_k^{-1}-w_k)w_k,
\label{eq1}
\ee
where $w_k=u_k/v_k$. 

Next, we rewrite this equation using the polarization angle $\gamma_k$ using
$u_k=\cos \theta_k \exp(i \gamma^1_{k})$ and 
$v_k=\sin \theta_k \exp(i \gamma^2_{k})$ under the condition of $|u_k|^2+|v_k|^2=1$. 
This leads to $\Delta(t)=
-\sum_p \sin (2\theta_p) \cos \gamma_p$, $\gamma_p=\gamma_p^1-\gamma_p^2$.
One can readily show from Eq.~(\ref{eq1}) that
\begin{widetext}
\ba
\dot \gamma_k w_k -i \frac{2}{\sin (2\theta_k)}\dot \theta_k w_k &=& 2|h_k| w_k -
2 w_k \lambda \sum_p \sin (2\theta_p) \cot (2\theta_k)  \cos \gamma_p \cos\gamma_k\nonumber
+2 i w_k \lambda \sum_p \frac{\sin (2\theta_p)}{\sin (2 \theta_k)}\cos\gamma_p\sin\gamma_k \\
&+& 2 m_0 w_k [- \cot (2\theta_k) \cos \gamma_k +i \frac{  \sin \gamma_k}{\sin (2 \theta_k)}] 
\ea
\end{widetext}
This further simplifies to two coupled first-order differential equations in $\theta_k$ and $\gamma_k$ 
given by 
\ba
\dot \gamma_k &=& 2|h_k|+2\lambda\sum_p \sin( 2\theta_p) \cot (2\theta_k) \times  
\big[ \cos (\gamma_p-\gamma_k) \non \\
&+&\cos (\gamma_p+\gamma_k) \bigr]-2m_0 \cot (2\theta_k) \cos \gamma_k 
\label{eq_gamma}
\ea
and 
\be 
\dot \theta_k = (- m_0  +\lambda\sum_p  \sin (2\theta_p) \cos \gamma_p)\sin \gamma_k
\label{eq_theta}
\ee

On the other hand, the generalized Kuramoto model~ \ct{kuramoto75, kuramoto84, strogatz88, bonilla05, rodrigues16} 
is represented by 
\be
\dot \eta_i = \omega_i + \sum_{j=1}^N M_{ij} \sin(\eta_i -\eta_j) + \epsilon_i(t)+ F\sin(\sigma t -\eta_i) 
\label{eq_kuramoto_forced}
\ee
which governs the synchronization between $N$ coupled (with coupling parameter $M_{ij}$) 
phase oscillators with phases $\eta_i$, 
oscillating  with individual frequency
$\omega_i$. The $F$ term represents the forcing strength. 
$\epsilon_i$ is a noise term that represents coupling to fast random degrees of freedom 
that leads to dissipation. 

Comparing Eq.~(\ref{eq_gamma}) and Eq.~(\ref{eq_kuramoto_forced}), one can infer that 
$\gamma_k$ plays the role of  $\eta_i$ in the Kuramoto model, 
and $2|h_k|$ of $\omega_i$. 
From our numerical results,  $\cos (\gamma_p+\gamma_k) $ in Eq.~(\ref{eq_gamma}) 
oscillates very fast and mimics the noise term $\epsilon_i$. 
When the laser field is on ($m_0\ne 0$), the last term of Eq.~(\ref{eq_gamma}) 
tries to pin the phases, and thus, corresponds to the forcing term in the Kuramoto model 
with $\sigma=0$. 
The coupling 
$M_{ij}$ is related to the $\lambda \sin (2 \theta_p) \cot(2 \theta_k)$
factor. 
This tells us two things. 
First, from the interaction dependence of the transition as shown in Fig.2 a in the main text, we notice that $\lambda(=1/2
(U/2-3V))$ is the key resource for connecting different $k$
modes leading to the DST. 
Second, the memory effect as shown in the inset of Fig.1 b (main text) can be understood by
the $\sin (2 \theta_p) \cot(2 \theta_k)$-term. If we switch off the field when this term is small, 
the DST transition may not occur even when $\lambda$ is sufficiently large. 

The synchronization order parameter for the Kuramoto model defined by  
\be
r\exp(i\psi)=1/N \sum_j \exp(i \eta_j)
\label{phase_trans}
\ee
is evaluated numerically from our data in momentum space $k$
by 
$r=\sqrt{(\sum_k \sin \gamma_k)^2+(\sum_k \cos \gamma_k)^2}/N$ and $\psi={\rm Arctan}(\sum_k \sin \gamma_k/\sum_k \cos \gamma_k)$.

We note that the mean field equations Eq.~(\ref{eq_gamma}) and 
Eq.~(\ref{eq_theta}) can also be expressed in terms of the momentum resolved CDW $\Delta_k$, current 
$J_k$ and bond   order $K_k$ parameters.  The three dimensional vector field in the momentum BZ are
given by  $\{\Delta_{\bm k},J_{\bm k},
K_{\bm k}\}=\{-\sin2\theta_{\bm k} \cos\gamma_{\bm k},-\sin2\theta_{\bm k} \sin\gamma_{\bm k},
\cos(2\theta_{\bm k})\}$.  
With the new variables, $\tan\gamma_k=J_k/\Delta_k$, $\tan 2\theta_k= \sqrt{J_k^2+\Delta_k^2}/K_k$, 
0
we can show (neglecting the $\dot K_k$ term since $K_k$ evolves slowly) that the mean field equations can be
recasted into 
\ba
\dot \Delta_k&=&-h_k J_k-\frac{2}{\Delta_k^2+J_k^2}\bigl [m_0 J_k (K_k J_k +\frac{\Delta_k}{K_k}) \non \\
&+& \lambda \Delta_k J_k \sum_p \frac{\Delta_p}{N} (K_k+\frac{1}{K_k}) \bigr ] \\
\dot J_k &=&h_k \Delta_k +\frac{2}{\Delta_k^2+J_k^2} \bigl [ \lambda \sum_p \frac{\Delta_p}{N} (K_k \Delta_k^2 -\frac{J_k^2}{K_k}) \non \\
 &-& 2 m_0 J_k (\frac{J_k}{K_k}-K_k \Delta_k) \bigr ]
 \label{decoupled}
\ea 
This expression highlights the interplay between current and CDW order. 

Moreover, we stipulate that in the weak interacting limit ($\la \ll1$) our model may lead to chaos. 
This is because there are some literatures~ \ct{maistrenko04, maistrenko08b, childs08} 
studying the Kuramoto models that reports the existence
of the chaotic regime. We will report on this aspect elsewhere.

\subsection{Investigating DST critically}
\label{s1ss2}
\textbf{Effect of the van-Hove singularity:}
In order to  investigate DST more clearly, we shall consider two lattice models \textbf{(i)} graphene \textbf{(ii)} graphene with an additional $AB$ hopping terms, both in the honeycomb lattice geometry. 
The motivation behind considering this extra lattice model is that to identify the influence of 
van-Hove singularity in the DST. The additional off-diagonal hopping term $\sin(k_x-k_y)\cos(k_x+k_y)$ is able to reduce the van Hove singularity of the density of 
states as shown in the main text.

\begin{figure*}[t]
\begin{center}
\includegraphics[width=18.0cm]{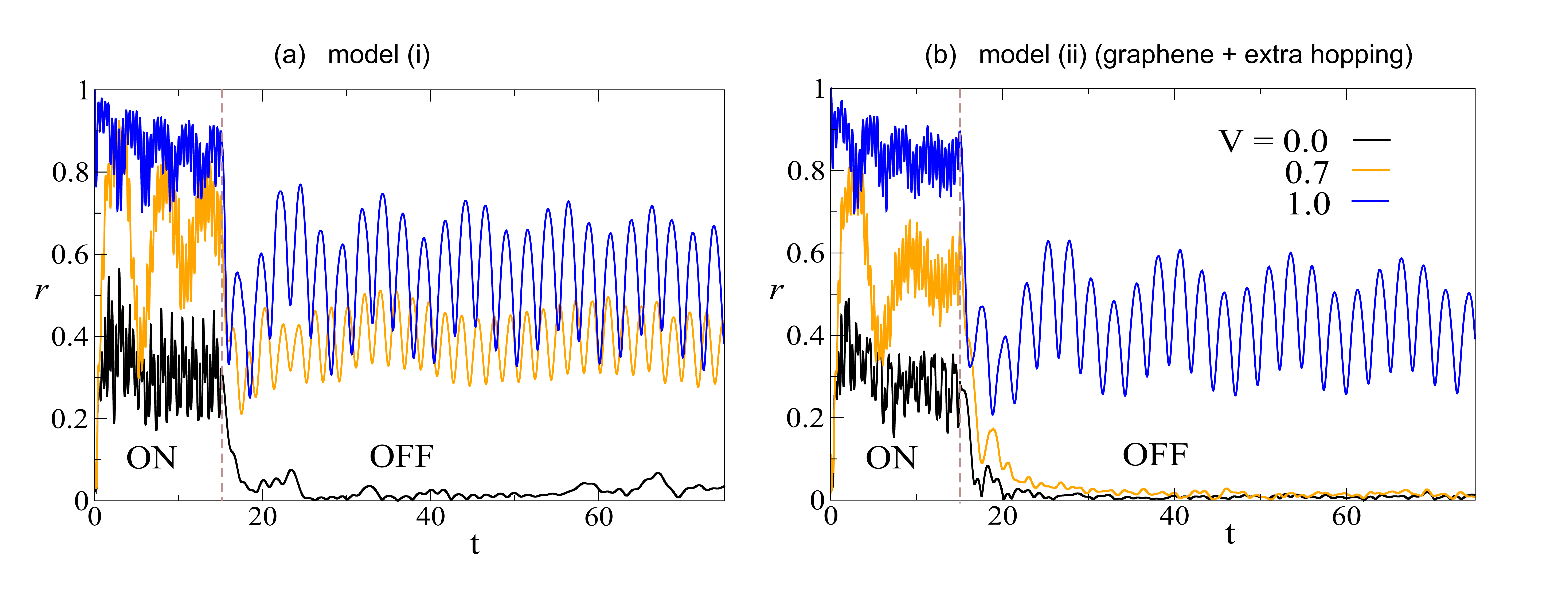}
\end{center}
\caption{Time evolution of synchronization order parameter $r$ for graphene, model (i) (a), and 
graphene with additional hopping i.e., model (ii) (b). We consider $\theta=\pi/10$ and $\omega=10$ and 
$N=100 \times 100$. 
}
\label{fig1}
\end{figure*}

We study the time evolution for 
synchronization order parameter $r$ for these above two models.  The common findings observed in both the models are the following: 
Starting from the non-ordered state ($V<V_{c}$), the order parameter $r$ starts from zero 
because the phase is indefinite, and 
increase quickly as the field activates the oscillation in the ON region. $r$ starts from unity if the groundstate is ordered. 
Prominent interaction dependence shows up in the switch OFF region. 
As the interaction is increased, the time averaged synchronization order parameter 
increases monotonically.  In the ON region, $r$ stays at a higher value compared to that in 
the OFF region. This reflects the fact that driving enhances the degree of synchonization and this 
is again related to the generation of more  electron-hole pairs during the driving.  Now, the 
marked difference between them is that for $V=0.7$, model (ii) with additional hopping  does not show any trace of  synchronization transition while model (i) already experiences the DST i.e., $r$
stays positive well above zero in the OFF region. Therefore, it is evident that the DST is not only depends on the external driving but also on the specific details of the model. The  interesting point is that upon the inclusion of $\sin(kx-ky)\cos(kx+ky) (c^\dagger_{A{\bm k}\sigma}c_{B{\bm k}\sigma}+c^\dagger_{B{\bm k}\sigma}c_{A{\bm k}\sigma})
$ term, the van-Hove singularity is reduced and the density of states (DOS) gets smoothened and broadened.
This has its impact on  $r$ in the switch OFF region where  $r$ continues staying in a much less values compared to model (i). 
This effect is clearly visible in Fig.~2 of the main text where 
$r_{\rm av}$ is plotted against $V$ to show the DST in a more generic way. 
 We use this data to calculate the average value of $r$, $r_{\rm av}$ as shown in the Fig.~2 of the main text.

The striking difference between these two models  
is that the jump in $r_{\rm av}$ for model (ii) while undergoing a DST is much more than that of the for model (i). 
Therefore, the singularity in DOS yields some trivial synchronization even in the non-synchronized phase where interaction is not able to initiate a collective phenomena.
This results in a relatively high value of $r_{\rm av}$ in the above phase. While in the synchronized phase, the interaction plays the key role over the trivial synchronization
factor. On the other hand, for model (ii), trivial synchronization is irrelevant and $V$ plays the pivotal role. As a result, difference between the values
of $r_{\rm av}$ in synchronized and non-synchronized phase is larger for model (ii) as compared to model (i).
Consequently, a clear sharp and singular upturn is observed for model (ii) in Fig. 2b of the main text. 


\begin{figure*}[thb]
\begin{center}
\includegraphics[width=15.0cm]{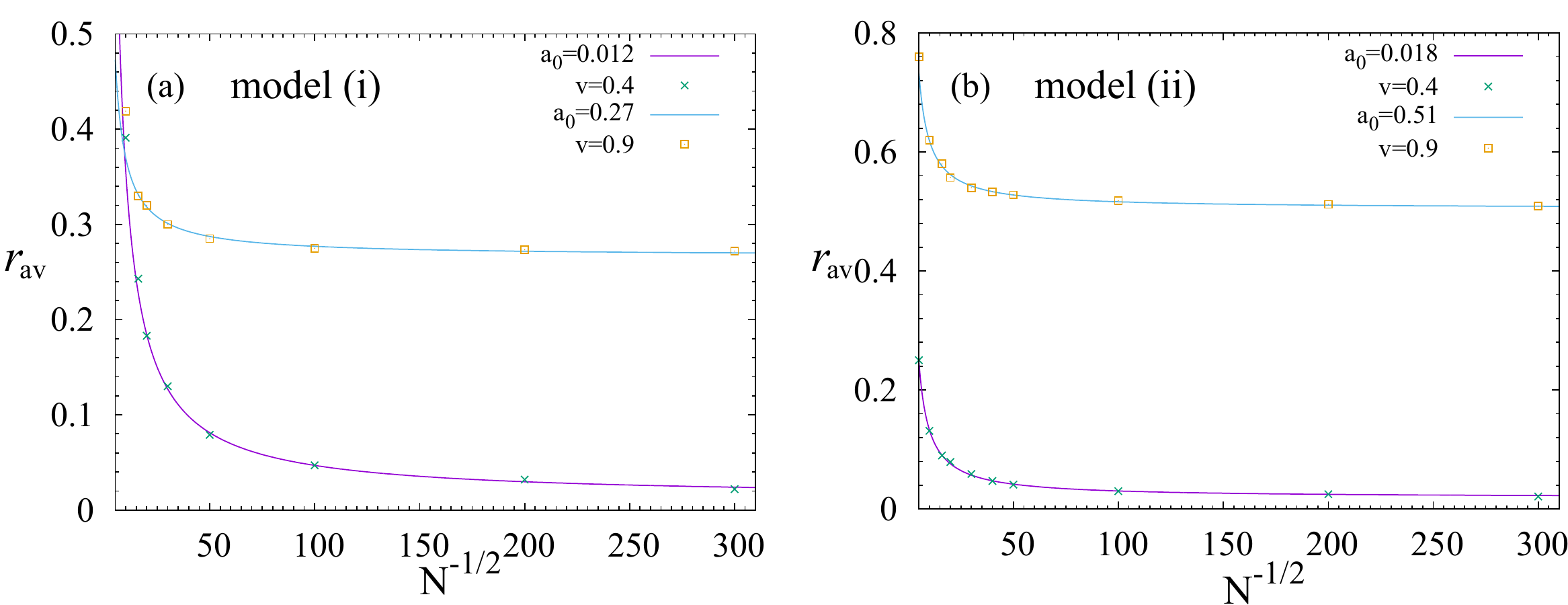}
\end{center}
\caption{(a) The plot shows that $r_{\rm av}$, obtained for model (i), follows a power law scaling: $r_{\rm av}=a_0+bN^{-1/2}$. Two distinct saturation 
values as captured by $a_0$ clearly suggests that there is synchronization transition.
Plot (b) is the same as (a) but for the model (ii).  
}
\label{fig1}
\end{figure*}

\textbf{Scaling analysis:}
Having described the DOS dependence on $r_{\rm av}$,  we shall now  numerically try to establish the dynamical synchronization transition in a more concrete way
by investigating the system size scaling properties of $r_{\rm av}$.
We first check the behavior 
for larger system size in honeycomb graphene lattice model (i). For that, we set $V=0.4$ in the non-synchronized regime  and $V=0.9$ in the synchronized regime 
with $\omega=10$. We obtain a power law dependence of $r_{av}$ on the system size 
\ba
r_{av}=a_0+bN^{-1/2}
\ea in both the 
regimes (see Figure \ref{fig1}).
The $N^{-1/2}$-scaling is the same as the Kuramoto model~\ct{bonilla05, rodrigues16}. 
 Interestingly, the $N\to \infty$ value of $r_{av}$ i.e., $a_0$ becomes very small ($O(10^{-2})$) for 
$V=0.4$ as compared to  $V=0.9$. For the synchronized regime, $a_0 \simeq 0.26$ which now conveys that
the synchronization transition is more clearly visible as we approach the thermodynamic limit. 
For validating our claim in a generic situation where the DOS is broadened and smoothened, we similarly investigate the model (ii). We find the 
same $N^{-1/2}$ scaling in both synchronized and non-synchronized regimes with $a_0\simeq 0.50$ and $O(10^{-2})$, respectively.

\begin{figure*}[thb]
\begin{center}
\includegraphics[width=15.0cm]{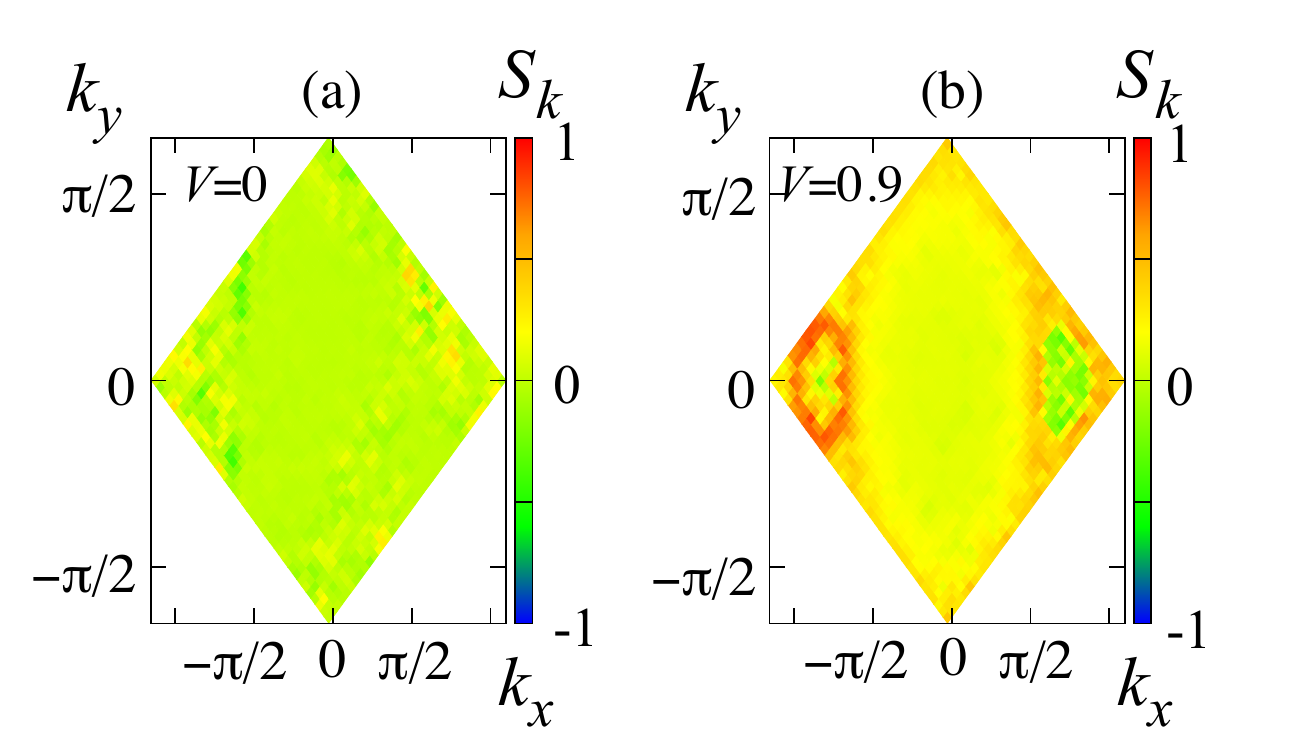}
\end{center}
\caption{The momentum resolved density plot of $S_{\bm k}$ for model (ii) are shown in (a) with $V=0$
and (b) with $V=0.9$. The non-interacting case does not exhibit any sign of correlation in the whole BZ while 
for the interacting case, a strong correlation is built up close to equi-energy surface enclosing Dirac point 
K.}
\label{fig:sync_new}
\end{figure*}

\textbf{Synchronization correlation for model (ii):}
We further study the synchronization correlation $S_{\bm k}$ for model (ii)
to investigate the location  of synchronized region the momentum space (see Fig.~\ref{fig:sync_new}).  
For the non-interacting case, the rotors (i.e., $\gamma_{\bm k}$), rotating randomly, are not able to generate any synchonized motion from any part of the momentum space
(see Fig.~\ref{fig:sync_new}~a). 
This is in contrast to the case with model (i) where some momentum modes near the van-Hove singularity actively
participate in the synchronization leading to an enhancement in $r_{av}$ even in the non-synchronized phase.  
The interacting case with $V=0.9$ for model (ii) is shown in Fig.~\ref{fig:sync_new}b. We find a
strong synchronized region (red patch) around the left Dirac point $K$ and 
in the energy space, this corresponds to
the $\varepsilon\sim \pm 1.5$ line in DOS for model (ii) (as shown in the inset of Fig. 2 b in the main text ). This refers to the fact that even when the singularity is absent, a peak in the density of states 
is enough to trigger the DST. 
We note that the K (left) and K' (right) points act differently in the $S_{\bm k}$-plot. 
This is because the bi-circular laser field acts in-equivalently to the two Dirac cones as explained in supplementary II.

\section{Floquet dynamical setup}
\label{sec:Floquet}

\subsection{Quasi-energy}
\label{sec:Floquet1}
Let us explain the Floquet quasi-energy spectrum \ct{shirley65, griffoni98, oka09, kitagawa11, lindner11a} of our model 
in the presence of bi-circularly polarized laser field. 
The Floquet Hamiltonian acting on the Fourier components of the Floquet state takes the form 
\begin{eqnarray} 
 \bf{H} &=& \left(
 \begin{array}{ccccc}  \ddots & \vdots  &  \vdots  & \vdots   &   \\ \cdots & \hat{H}_{0}-\omega & \hat{H}_{1} && 
 \\    &  \hat{H}_{-1} & \hat{H}_{0} & \hat{H}_{1} &
  \\  & & \hat{H}_{-1} & \hat{H}_{0}+\omega &  \cdots \\  & \vdots  &  \vdots  & \vdots   &  \ddots
 \end{array} \right) \nonumber ,
 \end{eqnarray} 
 with $\hat{H}_{n} = \frac{1}{T}\int^{T}_{0} \hat{H}_k(t) e^{it\omega n} dt$. Here, $H_k(t)$ is the time dependent lattice Hamiltonian 
 as given in Eq.~(2) of the main text.

\begin{figure*}[bht]
\begin{center}
\includegraphics[width=14.0cm]{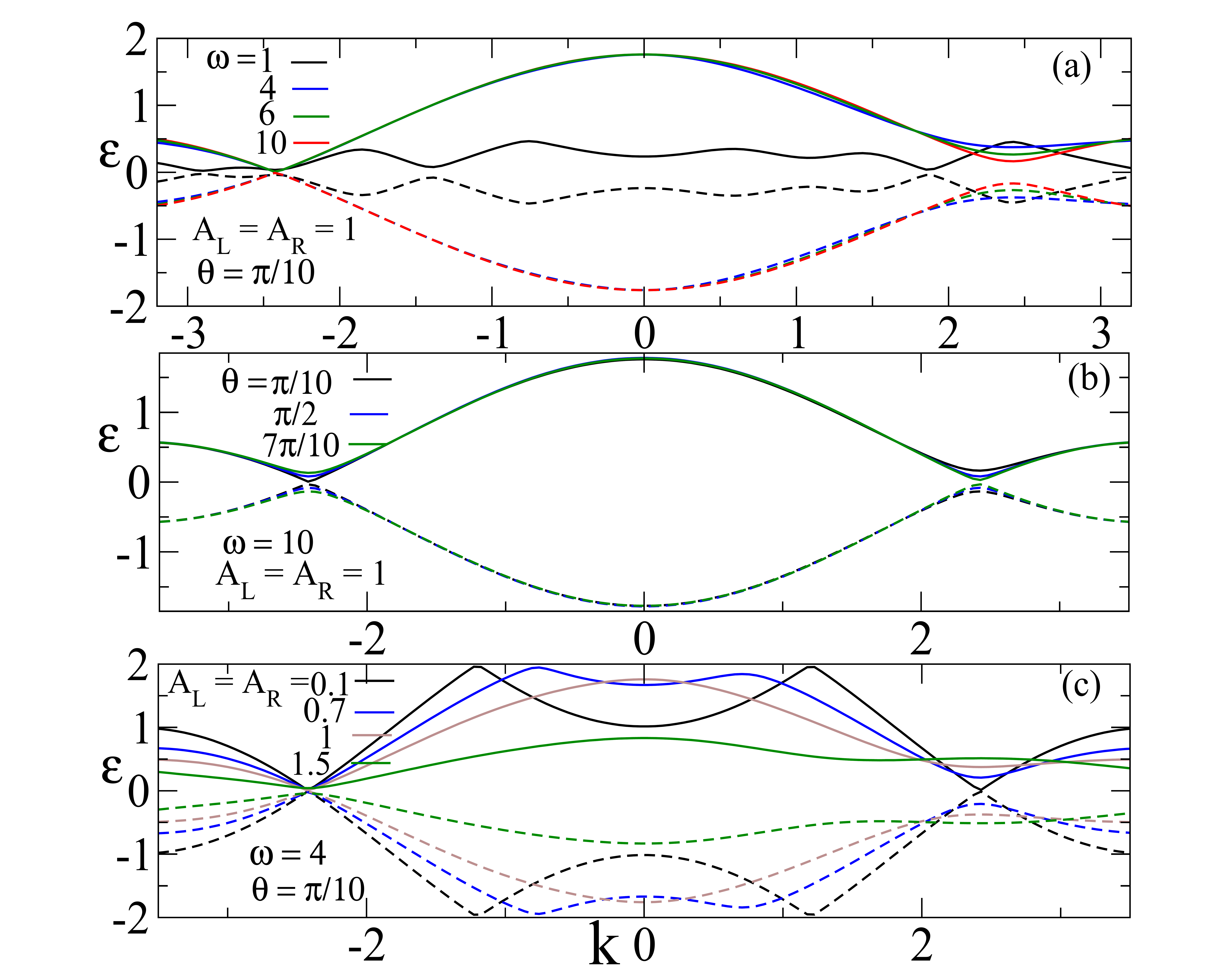} 
\end{center}
\caption{ Variation of quasi-energy $\varepsilon$ with $k_x$ keeping $k_y=0$ for different values of 
frequency (a), relative phase (b), and amplitude (c). (a): The gap at  $(-K,0)$ is much smaller than 
than of the at $(K,0)$. This asymmetry of gaps is reduced as one increases $\omega$ i.e., the gap appearing at
$(K,0)$ decreases, whereas, the other one changes insignificantly. The gap also increases in the central part with 
$\omega$ within the same temporal BZ. (b): Gap at $(-K,0)$ increases with increasing $\theta$ within the range $0<\theta< \pi$, while gap at  $(K,0)$ decreases.
For $\theta=\pi/2$ both of them become symmetrically gapped.  (c): Both the Dirac points are insignificantly 
gapped for smaller amplitude; the gap at $(K,0)$ increases with $A_L$. Although, for smaller amplitude 
the gaps between two consecutive photon sector appear symmetrically around $k_x=0$; they disappear for higher 
amplitudes.
} \label{fig_qe} \end{figure*} 

We investigate the behavior of quasi-energy as function of $k_x$ with $k_y=0$ as shown in Fig.~(\ref{fig_qe}).
The gap at K$=(K,0)$ can be tuned by varying the frequency and amplitude. The gap at the other Dirac point K'$=(-K,0)$ does 
not change significantly. The two quasi-energies of two states residing in the same photon sector or in two consecutive 
time BZs come closure for low amplitude and low frequency. The relative phase $\theta$ is able to tune 
both the gaps appearing at the Dirac points. The gaps at $(-K,0)$ point change in an opposite way with $\theta$.
These points are identically gapped for $\theta=\pi/2$.

\subsection{Brillouin-Wigner evaluation of the Floquet Hamiltonian}
\label{sec:Floquet2}
We now formulate the high frequency effective Hamiltonian which enables us to study the phase diagram of Chern 
numbers in the various parameter space.  To this end, we employ the Brillouin-Wigner (BW) perturbation theory to construct the high frequency 
effective Hamiltonian on the projected zero-photon subspace of Floquet theory \ct{mikami16}. This correctly reproduces the quasienergies and the 
eigenstates associated with the original Floquet Hamiltonian up to order $1/\omega$. Specfically,
a projector operator projects the whole Hilbert space $H\otimes T$ onto the zero photon subspace $H\otimes T_0 \sim H$. On the 
other hand, the wave-operator restores the eigenstates of $H\otimes T$ from the projected space of $H\otimes T_0$. The effective 
Hamiltonian is then expressed in terms of this wave operator that can be determined in powers of $1/\omega$ recursively.  This theory 
has a number of advantages over the other high frequency perturbation theories namely, i.e the Floquet-Magnus expansion and 
van Vleck expansion, being that the BW theory rectifies the driving phase dependence problem appearing in the Floquet-Magnus expansion.
Additionally, BW theory can be used to get the higher order terms which are challenging to compute using the van Vleck theory.

Having stated the BW theory in a nutshell, we revert to the associated  final effective Hamiltonian
 $H_{\rm BW}
=\sum_{n=0}^\infty H_{\rm BW}^{(n)}$ with $$H_{\rm BW}^{(0)}=H_{0,0}, 
 ~~H_{\rm BW}^{(1)}=
\sum_{\{n_{i}\}\neq0}
\frac{H_{0,n_{1}}H_{n_{1},0}}{n_{1}\omega}$$
 and $$H_{\rm BW}^{(2)}=
\sum_{\{n_{i}\}\neq0}
\left(
\frac{H_{0,n_{1}}H_{n_{1},n_{2}}H_{n_{2},0}}{n_{1}n_{2}\omega^{2}}-\frac{H_{0,n_{1}}H_{n_{1},0}H_{0,0}}{n_{1}^{2}\omega^{2}}
\right).$$
In the process of quantifying each term  $H_{m,n}$ in the above expression, we need to compute the hopping strength accordingly in the 
presence of the laser fields \ct{mikami16}. 
The hopping strength from $A$ sub lattice to $B$ sub lattice 
becomes $t_{AB}^{n,l}=\sum_m J_{-n-2m}(-A_R)J_m(A_L)e^{-i m \theta} e^{\frac{i 2\pi l
  (3m +n)}{3}}$. Similarly,
  the $B$-$A$ hopping is given by $t_{BA}^{n,l}=\sum_m J_{-n-2m}(A_R)J_m(-A_L)e^{-i m \theta} e^{\frac{i 2\pi l
  (3m +n)}{3}}$. While doing the calculation of the perturbation theory term by term, a variety of new hopping terms 
  gets generated. Below shall explicitly mention them.

\begin{figure*}[t]
\begin{center}
\includegraphics[width=15.0cm]{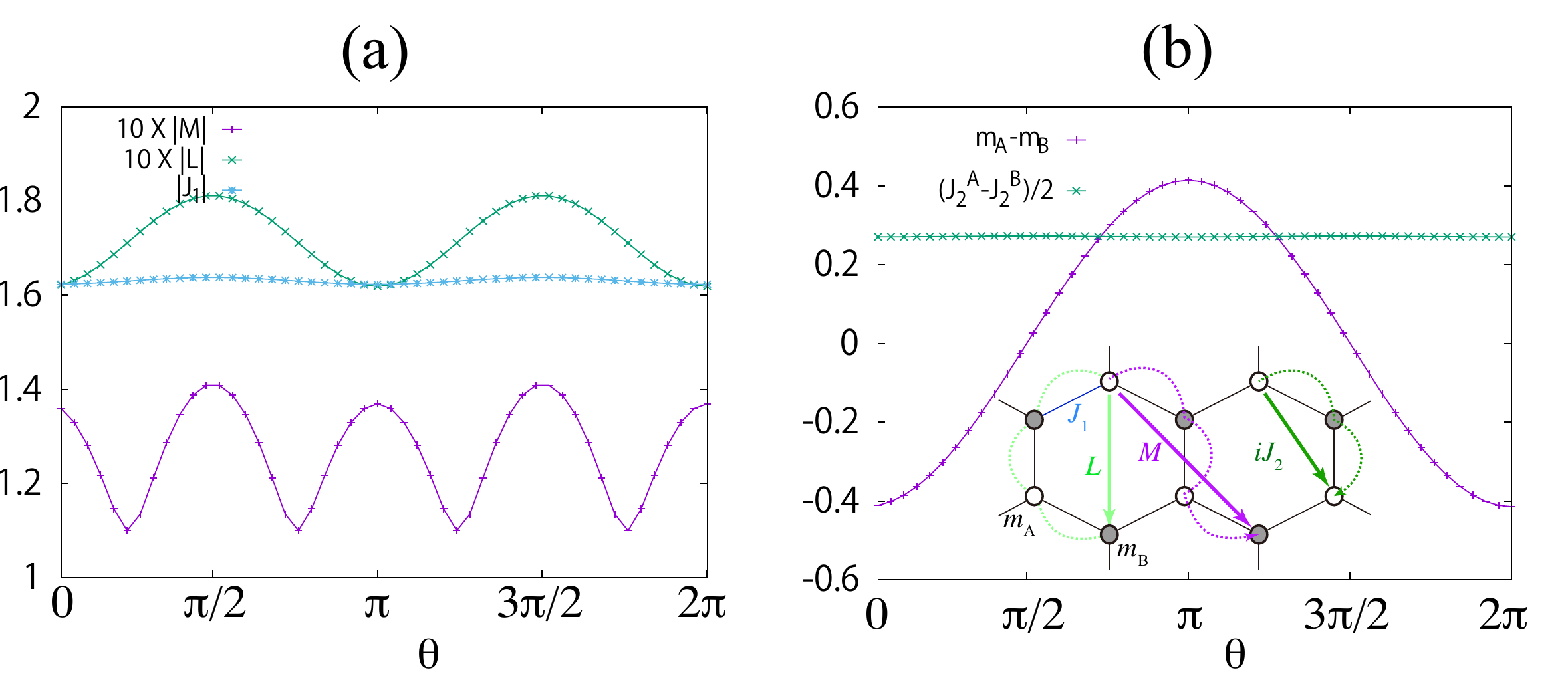}
\end{center}
\caption{(a) The plot shows the variation of the magnitude of the off-diagonal hopping terms 
with $\theta$. As expected, $|J_1|$ is most significant contributor and $|M|$ contributes 
insignificantly. The common fact is that $|J_1|$ and $|L|$ both exhibits a maxima near 
$\theta=\pi/2$ and a minima at $\theta=\pi$. While $|M|$ shows local maxima at $\theta=\pi/2$, $\pi$ and minima at $\theta=3\pi/4$. (b) The plot shows the diagonal mass terms and the 
hopping term to same  sub-lattice as a function of $\theta$. Interestingly, $m_A-m_B$ changes 
its sign when $\theta=\pi/2$. Inset depicts the geometrical description of different terms.}
\label{fig:BW_hopping}
\end{figure*}  

We now explicitly write the NN hopping $J_1$ obtained from $H_{\rm BW}^{(0)}$, $H_{\rm BW}^{(2)}$ terms
\begin{widetext}
\ba
a_i^{\dagger}b_{i+p}: J_1=t_{AB}^{0,p} &+&\sum_{n\neq 0}\frac{-1}{n^2\omega^2} \left( \sum_{q=0,1,2} t_{AB}^{-n,p}t_{BA}^{n,q} t_{AB}^{0,q}
\sum_{q=0,1,2}^{q\neq p} t_{AB}^{-n,q}t_{BA}^{n,q} t_{AB}^{0,p}\right) \non \\
&+&\sum_{n,m\neq0}\frac{1}{mn\omega^2}
\left( \sum_{q=0,1,2} t_{AB}^{-n,p}t_{BA}^{n-m,q} t_{AB}^{m,q}+\sum_{q=0,1,2}^{q\neq p} t_{AB}^{-n,q}t_{BA}^{n-m,q} t_{AB}^{m,p} \right)
\label{eq_N1}
\ea
\end{widetext}
where $p$ represents the links $l=0,1,2$ for NN hopping. 
Now,  NNN terms $J_2$, obtained from $H_{\rm BW}^{(1)}$, are given by
\ba
a_i^{\dagger}a_{i+p-q}&:& J^{A}_2=\sum_{n\neq 0}\frac{-1}{n\omega}  t_{AB}^{-n,p}t_{BA}^{n,q}\\ 
b_i^{\dagger}b_{i+p-q}&:& J^{B}_2=\sum_{n\neq 0}\frac{-1}{n\omega}  t_{BA}^{-n,p}t_{AB}^{n,q}.
\label{eq_N2}
\ea
The mass terms $m_A$ and $m_B$ can be obtained from the above expressions of $a_i^{\dagger}a_{i+p-q}$ and $b_i^{\dagger}b_{i+p-q}$
with $p=q$.
The third nearest neighbour hopping connected by $M$-links \ct{mikami16} $J_3$, obtained from $H_{\rm BW}^{(2)}$,  are given by 
\ba
a_i^{\dagger}b_{i+2p-q}:  &J_3&=\sum_{n\neq 0}\frac{-1}{n^2\omega^2} [t_{AB}^{-n,p}t_{BA}^{n,q} t_{AB}^{0,p}] \non \\
&+&\sum_{n,m\neq 0}\frac{1}{n m\omega^2} t_{AB}^{-n,p}t_{BA}^{n-m,q} t_{AB}^{m,p},
\label{eq_m_link}
\ea
where $p\neq q= 0,1,2$.
The other type of third nearest neighbour hopping connected by $L$-links $J_4$ \ct{mikami16}, obtained from $H_{\rm BW}^{(2)}$,  are given by 
\ba
a_i^{\dagger}b_{i+p-q+r}: &J_4&=\sum_{n\neq 0} \frac{-1}{n^2\omega^2} [t_{AB}^{-n,p}t_{BA}^{n,q} t_{AB}^{0,r}] \non \\
&+&\sum_{n,m\neq 0}\frac{1}{n m\omega^2} t_{AB}^{-n,p}t_{BA}^{n-m,q} t_{AB}^{m,r},
\label{eq_l_link}
\ea
with $p\neq q \neq r= 0,1,2$.
The hermitian conjugate of the above terms (Eq.(\ref{eq_N1}), (\ref{eq_m_link}), (\ref{eq_l_link})) gives 
the hopping from $B$ sub lattice to $A$ sub lattice; in this case $t_{AB}$ is replaced by $t_{BA}$ and vice versa.
The final Hamiltonian looks like 
\ba
H_{BW}&=&\sum_{i,p}^{NN}J_1 a_i^{\dagger}b_{i+p}+\sum_{i,p\neq q}^{NNN,A(B)}J^{A(B)}_2 a(b)_i^{\dagger}a(b)_{i+p-q}  \non \\
&+& \sum_{i,p\neq q \neq r}^{L-links}J_4 a_i^{\dagger}b_{i+p-q+r} +  \sum_{i,p\neq q}^{M-links}J_3 a_i^{\dagger}b_{i+2p-q} +h.c \non  \\
&+& \sum_{i}^{A(B)-sub lattice} m_{A(B)} a(b)_i^{\dagger}a(b)_{i}  + {\rm h.c.}
\label{eq_final}
\ea
Hence, the high frequency expansion gives rise to a momentum independent mass term that breaks the inversion 
symmetry. This is the reason why CDW order oscillates with real time inside the switch OFF region.  
Having formulated the Hamiltonian in real space, one can go to Fourier space where $H_{BW}$ is decomposed
in different 
momentum segments, $H_{BW}=\prod_k H_{BW}^k$.  The mass terms associated with $\sigma_z$ is responsible 
for opening up a gap in the spectrum which eventually leads to non-trivial topological phases.

\begin{figure*}[ht]
\begin{center}
\includegraphics[width=8.cm]{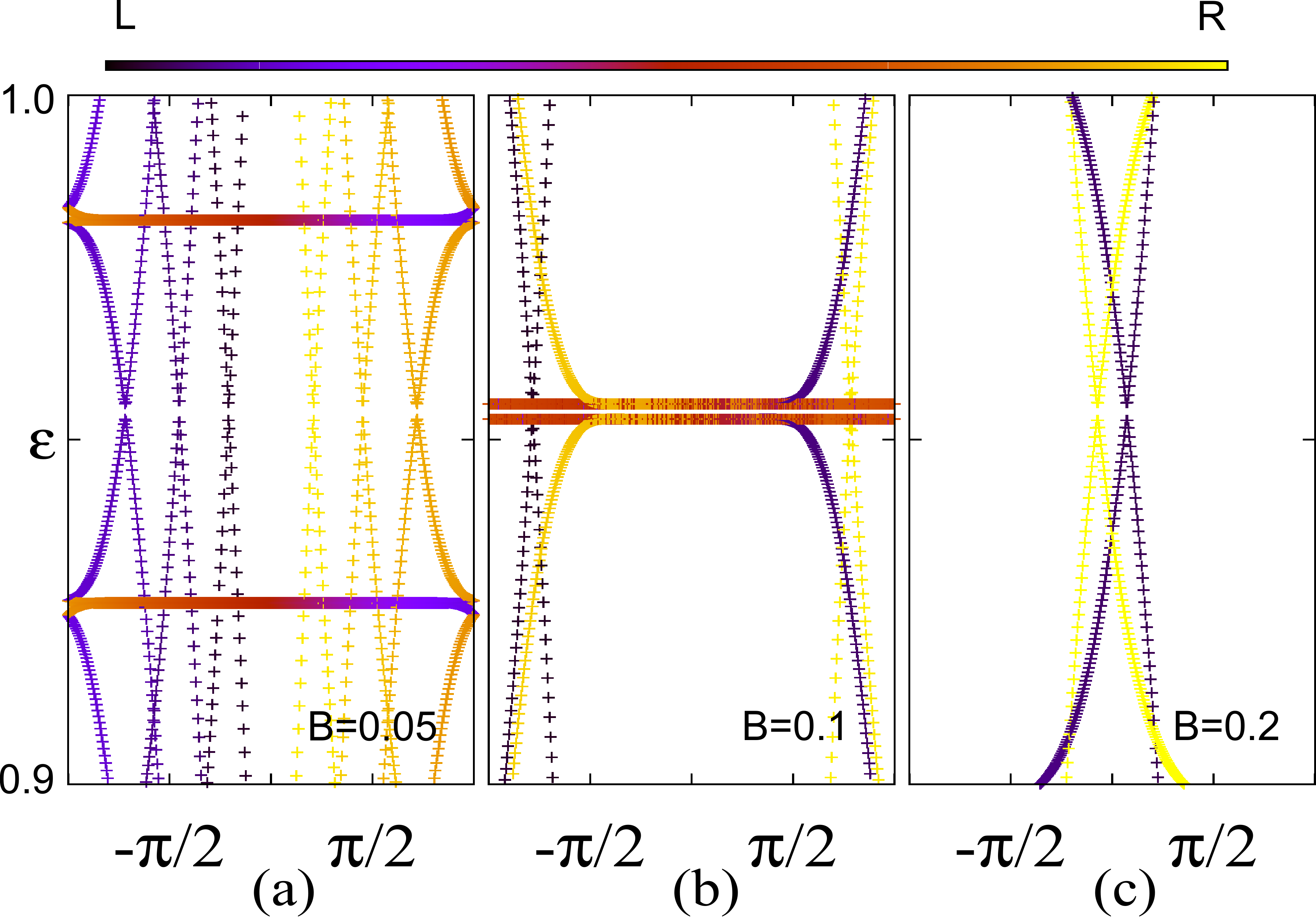} 
\includegraphics[width=8.cm]{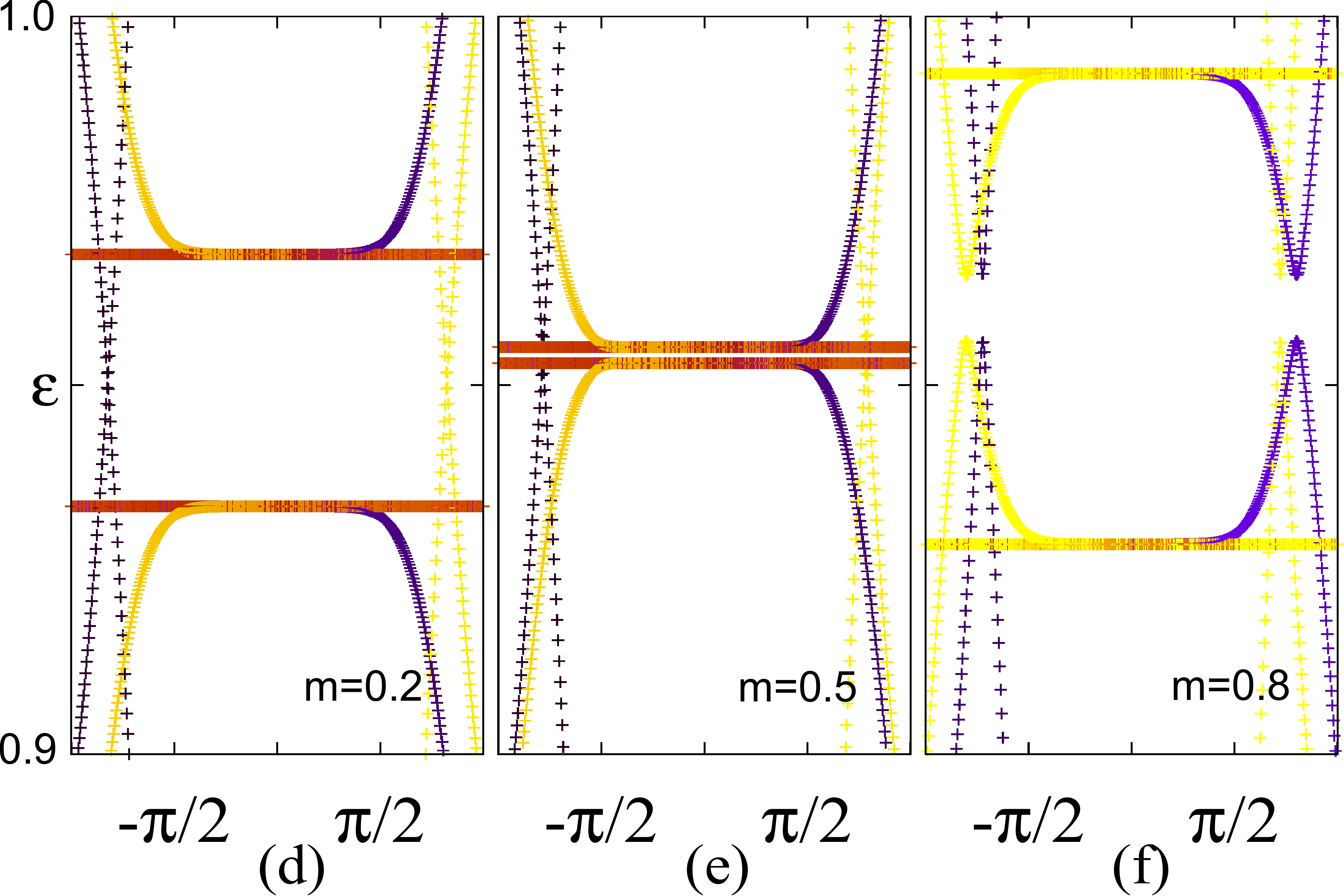} 
\end{center}
\caption{
Quasi-energy dispersion of the Floquet Hamiltonian as function of $k_y$ for various parameter
values. The colour map indicates the weight of states on each lattice sites. The first three panels show that
 increasing $B$  leads to a topological 
phase transition between two distinct topological phase by a closing of the bulk gap; here
$m$  is kept at $0.5$. (a) Topological phase 
with three pairs of edge state. (b) Closing of the bulk band gap due to the vanishing the gaps between of the flat bulk modes. (c) Another topological phase emergences with one pair of edge state. The
last three panels show the transition from a topological phase to a non-topological phase by varying $m$ keeping $B=0.1$.
(d) A situation where only 1 pair of edge state appear between the bulk flat band. (e) Bulk gap closing.
(f) Non-topological phase having no $\pi$ edge states. In all case the lattice size $N=200$.}
 \label{fig6}
 \end{figure*}

One can think of $H_{BW}^k$ as a spin Hamiltonian  with ${\bm k}$ dependent magnetic field terms i.e., $H_{BW}^k=\mu B_{\bm k}\cdot \sigma$. Since the 
$2 \times 2$, momentum space effective Hamiltonian consists of the different kinds of hopping terms e.g., nearest neighbour hopping $J_1$, long 
range hopping $J_2$, $J_3$ and $J_4$ and on-site terms $m_{A(B)}$. Among these terms $J_2$ represents the hopping between the same sub-lattice. Therefore, 
the diagonal terms are given by momentum independent mass terms along with the momentum dependent $J_2 \cos ({\bm k} (p-q))$. On the other hand, the off-diagonal 
terms are given by $J_1 \exp(\pm i {\bm k} p)$, $J_3 \exp(\pm i {\bm k}(2p-q))$ and $J_4\exp(\pm i {\bm k}(p-q+r))$. Therefore, the explicit momentum dependent 
magnetic fields are $B_{{\bm k}x}= J_1 \cos ({\bm k}p) + J_3 \cos ({\bm k}(2p-q)) + J_4 \cos ({\bm k}(p-q+r))$, 
$B_{{\bm k}y}= J_1 \sin ({\bm k}p) + J_3 \sin ({\bm k}(2p-q)) + J_4 \sin ({\bm k}(p-q+r))$
$B_{{\bm k}z}=J_2 \cos ({\bm k}(p-q)) + (m_A-m_B)$. In order to clearly investigate the effect of each terms, we study the momentum independent 
coefficients namely, $J_1$, $J_2$, $J_3$, $J_4$, and $m_A-m_B$ as a function of $\theta$. This would provide a clear connection to the spin echo analogy for 
$\theta=\pi/2$ as discussed in the main text.

Figure~\ref{fig:BW_hopping} shows the behavior of the effective terms
while  $\theta$ is varied. Since $J_1$ represents the nearest neighbour hopping, its magnitude becomes largest among the hoppings. The $M$-link hopping refers to the maximum  long range hopping amplitude and hence its magnitude is the smallest. While $L$-link hopping are intermediate range. The interesting common fact about all the above hopping is that 
they all show oscillatory behavior with $\theta$ and acquire maximum magnitude at $\theta=\pi/2$, $3\pi/2$. On the other hand, the NN-link hopping and 
$L$-link hopping both acquires  minimum magnitude at $\theta=0$. The $M$-links hopping becomes minimum for $\theta=\pi/4$, $3\pi/4$, $5 \pi/4$  and $7\pi/4$. It  again shows a secondary maxima at $\theta=0, \pi$ and $2\pi$.  On the other hand, for diagonal mass term $m_A-m_B$ changes its sign at $\theta=\pi/2$ and $3\pi/2$. It becomes maximally negative and positive at $\theta=0$ and $\pi$. 
Therefore, it can be inferred that  $\theta=0$ and $\pi$ refers to the two extreme situation where 
CDW order acquires its maximum magnitude and the off-diagonal terms become insignificant. 
On the other hand, 
for $\theta=\pi/2$ and $3\pi/2$, the mass term becomes zero and gives room for the other terms 
to significantly alter the dynamics of the CDW order. 

Relating the above finding to the spin echo technique, employed in NMR, one can say that 
$z$-component magnetization is caused by $B_{{\bm k}z}$. 
When the ``transverse fields" $B_{{\bm k}x}$, $B_{{\bm k}y}$ acquire maximum values, the magnetization along $z$-direction can get flipped. 
This is due to fact that $\sigma_{x,y}$ does not commute with $\sigma_z$ and hence this off-diagonal term results in a precession. 
We note that the above argument is based on the Floquet effective Hamiltonian in the non-interacting case. 
For the interacting case, this argument is not guaranteed to hold. 
However, from our numerical findings, it still provides a qualitative understanding for the flipping of CDW order as displayed in Fig. 1e (main text).

\section{Time glide symmetry, topology and Road to self-maintained Floquet crystalline states}

We now investigate the emergence of edge state in a minimal model, considered for the 
switch off region, with an open boundary condition in $x$-direction. Existence of such edge 
states that persist and appear upon band gap closings, directly conveys the topological nature of
the model in certain parameter regimes. This indicates the potential of DST setups to create novel Floquet
crystalline states that are protected by space-time symmetries.

In the OFF region, when DST takes place, an oscillating CDW 
order  is formed. Let us investigate  this situation using a simplified lattice model
\begin{eqnarray}
 H(t)&=&    H_0+\lambda \Delta (t)\sum_i(c^\dagger_{Ai} c_{Ai}-c^\dagger_{Bi} c_{Bi})
\label{eq_ham}
\end{eqnarray}
where $H_0$ denotes the non-interacting tight binding model for electrons on the honeycomb lattice as given in Eq.~(2) in the main text,
and $\Delta(t)=m\sin (\Omega_0 t)$  represents the CDW order 
oscillation and $\Omega_0=1.9$ is the system selected frequency.  
Importantly, this model has a
chiral symmetry: $\Gamma \mathcal{H}_{\bm k}(t) \Gamma=-\mathcal{H}_{\bm k}(-t)$ 
in the momentum space picture with ${\cal H}_{\bm k}(t)=\Delta(t) \sigma_z +h_x \sigma_x +h_y \sigma_y$ and  $\Gamma=\sigma_z$. 
The corresponding real space model has a time glide symmetry as alluded  in the main text,  $H(t)=MH(t+T/2)M^{-1}$ ($T=2\pi/\Omega_0$)
where $M$ is the mirror symmetry  that is broken by the AB-sublattice potential. 
The Floquet topological crystalline states protected by the time glide symmetry can now directly be investigated by checking whether the
system supports $\pi$ chiral edge states. Their existence is also closely tied to the chiral symmetry and have been proposed in theoretical models before \cite{Tg1}.
In order to look for edge states, we consider an extra static magnetic field $B=(0,0,B)$. This will modify the hopping through 
 a phase factor $\exp(i \int A.dr)$ with vector potential $A=(0,Bx,0)$. Given this model, 
we consider an armchair edge in the brick wall lattice geometry.  We diagonalize the Floquet Hamiltonian to obtain 
the quasi-energy, quasi-states and the associated average weight of each state on the lattice sites.

The results plotted in Fig.~(\ref{fig6}) show that, depending on the parameter regions, $\pi$ edge mode indeed appear. 
The $\pi$ edge modes are states spatially localised at the edge of the system appearing at $\varepsilon=\pm \Omega_0/2$
in the Floquet quasi-energy space, and characterise dynamically extended topology of periodically driven systems~\cite{Rudnerprx}. 
We also find several topological transitions when the bulk gap vanishes and reopens. The left three panels of Fig.~(\ref{fig6}) convey that there is a
topological transition between two topological phases as the field strength is increased 
as the gap close and reopens at
$B=0.1$ while the strength of the CDW oscillation $m$ is kept fixed to $m=0.5$. We also remark that  the topological phases
are always characterised by an even number of edges modes i.e., edge states appear always in pairs. For example, 
Fig.~(\ref{fig6}a) depicts a situation where three pairs of modes appear at quasi-energy $\varepsilon=\Omega_0/2$
at the left boundary and another three pairs of modes at the right boundary; each pair consists of 
two opposite chirality. Moreover, it is noteworthy that bulk bands close to  $\varepsilon=\Omega_0/2$
becomes flat and the band gap vanishes at the topological transition point.  
One can also vary $m$ to encounter a topological phase transition as 
observed in Fig.~(\ref{fig6}e) with $B=0.1$. In the right three panels of Fig.~(\ref{fig6}), we exemplify a situation 
where a topological phase is separated from a non-topological phase by a bulk gap closing. The non-topological 
phase does not contain any edge state at $\varepsilon=\Omega_0/2$. This asserts that topological phases protected by Floquet crystalline symmetries can indeed be induced by the DST setup.



\begin{figure*}[ht]
\begin{center}
\includegraphics[width=15.cm]{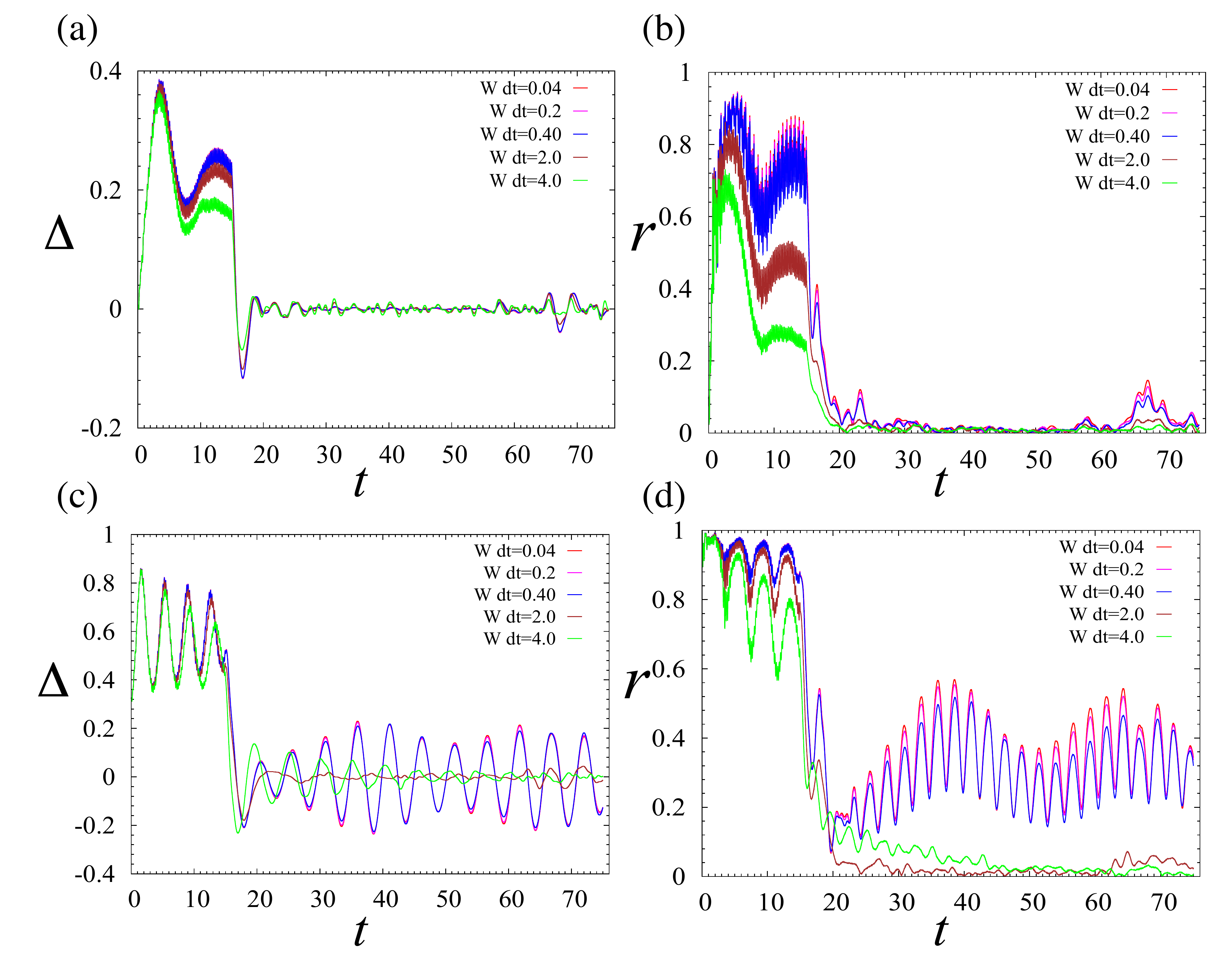} 
\end{center}
\caption{The time evolution of $\Delta$ and $r$ are shown for $V=0.3$ in (a) and (b),
respectively. The same for $V=0.8$ are shown in (c) and (d). Here, we consider graphene model
(i) with momentum-time dependent random mass term. The strength of this disorder 
is given by $Wdt$. The synchronization is substantially affected by the disorder as observed for 
$V=0.8$. Here, $\omega=10$ and $\theta=\pi/10$.}
 \label{fig:random_mass_sync}
 \end{figure*}


\section{Going beyond mean field}
\label{s4}

We did not consider heating in our calculation. 
It has been a central topic of 
interest to study the heating effect in a periodically driven system. Recent studies in the context of many body systems
show that  prethermalization takes place in the high-frequency regime~\cite{PhysRevB.95.014112,Weidinger:tb,PhysRevLett.116.120401}. 
What is argued there is that for a finite time window, the dynamics of an interacting system 
can be described within the Floquet effective Hamiltonian approach~\cite{PhysRevLett.116.120401}. 
In our problem, the pulse duration of the laser field is
finite and can be made short to fit within the Floquet prethermalization window.

We note that when the pulse duration becomes exponentially long, 
then the 
system would not be restricted inside the 
prethermal regime. Once the system starts absorbing energy, there would be substantial reduction of the synchronization. Hence 
the synchronization order parameter $r$ is expected to decrease. Similarly, in the switch off regime, $r$ would 
decrease. 
In order to clarify the heating effect arising from scattering, we perform an additional numerical calculation below.

In the mean field approximation no scattering process are considered. 
However, in order to accommodate the effect of relaxation due to electron-electron or electron-phonon scatterings, 
we extend the mean field treatment by considering a random mass term in the momentum space 
which changes its value for each 
momentum and time instant $M=\eta(t,k)$ with $\eta$ chosen from a Gaussian distribution centered around zero. 
This term mimics the effect of a Hubbard-Stratonovich field that arrises when 
an electron-electron interaction is incorporated~\cite{PhysRevB.79.035320}. 
The dynamical evolution of CDW oder $\Delta$ and the synchronization order parameter $r$ are shown in  Fig.~\ref{fig:random_mass_sync}.
For low disorder strength, $r$ is quantitatively unaltered in the both switch off and on region. The CDW order 
behaves similarly as $M=0$. For substantial disorder strength, $r$ decreases in the switch on region. While 
in the switch OFF region, $r$ decreases to zero within a window of time after the switch on region.
This window of time gradually decreases with increasing disorder strength.
Therefore, the long time average value of $r$, $r_{av}$
becomes zero when the disorder is substantial. 
This refers to the fact that the synchronization transition would disappear for strong 
disorder. 

Now, we shall elaborate the above discussion. 
The interesting observation in the OFF region is that for substantial disorder strength $W dt>0.4$, CDW order 
$\Delta$ continues to show coherent oscillation in a short window of time $15<t<40$
(see Fig.~\ref{fig:random_mass_sync}~c). Inside this temporal regime, 
$r$ also remains finite, signifying the fact that system remain in the synchronized phase 
(see Fig.~\ref{fig:random_mass_sync}~d). As time passes by $t>40$,
$\Delta$ starts dephasing and continues aperiodic random oscillation for further time. There, synchronization order 
parameter $r$ decreases and becomes small. Hence, under a sufficiently long relaxation time, we expect that  $r_{\rm av}$ as well as the 
oscillation amplitude in $\Delta$ diminish to zero.
The time duration for which the coherence is remained is a few hundred time periods: $\delta t=2 m \pi/ \omega$, with 
$m \sim O(10^2)$. The higher strength of the disorder is equivalent to a higher
degree of momentum mixing  due to the electron-electron scattering. 
For relatively small interaction strength 
$V=0.3$, when graphene remains in a non-synchronized phase with the same driving frequency, the 
temporal window $\delta t$ becomes vanishingly small, as expected (see Fig.~\ref{fig:random_mass_sync} a, b).

We can now make some general comment regarding the scattering phenomena. 
We know from the existing literature that relaxation time in graphene is large compared to the strongly 
correlated cuprates where it is femtosecond. Experimentally, it has been shown that relaxation time in graphene is of the 
order of picoseconds \cite{Gierz:tg}. Additionally, in low temperature, the phonon scattering is reduced. 
Hence, we think that even in the presence of scattering, a few cycles of coherent oscillation in $\Delta$ can be observed after switching off the 
laser field.

\bibliographystyle{naturemag}
\bibliography{DS}

\end{document}